\documentclass[journal=jacsat,manuscript=article]{achemso}
\usepackage{xr}
\makeatletter
\newcommand*{\addFileDependency}[1]{
  \typeout{(#1)}
  \@addtofilelist{#1}
  \IfFileExists{#1}{}{\typeout{No file #1.}}
}
\makeatother

\newcommand*{\myexternaldocument}[1]{
    \externaldocument{#1}
    \addFileDependency{#1.tex}
    \addFileDependency{#1.aux}
}
\myexternaldocument{Supportinginformation}

\usepackage[version=3]{mhchem} 
\usepackage[T1]{fontenc} 
\usepackage{gensymb}

\author{Shili Yan}
\affiliation[BAQIS]
{Beijing Academy of Quantum Information Sciences, 100193 Beijing, China}
\alsoaffiliation[EqualContribution]
{These authors contribute equally}

\author{Haitian Su}
\affiliation[PKU]
{Beijing Key Laboratory of Quantum Devices, Key Laboratory for the Physics and Chemistry of Nanodevices, and School of Electronics, Peking University, Beijing 100871, China}
\alsoaffiliation[ICMMP]
{Institute of Condensed Matter and Material Physics, School of Physics, Peking University, Beijing 100871, China}
\alsoaffiliation[EqualContribution]
{These authors contribute equally}

\author{Dong Pan}
\affiliation[CASIOS]
{State Key Laboratory of Superlattices and Microstructures, Institute of Semiconductors,Chinese Academy of Sciences, P.O. Box 912, Beijing 100083, China}
\alsoaffiliation[EqualContribution]
{These authors contribute equally}

\author{Weijie Li}
\affiliation[PKU]
{Beijing Key Laboratory of Quantum Devices, Key Laboratory for the Physics and Chemistry of Nanodevices, and School of Electronics, Peking University, Beijing 100871, China}
\alsoaffiliation[AAIS,PKU]
{Academy for Advanced Interdisciplinary Studies,Peking University,Beijing 100871, China}

\author{Zhaozheng Lyu}
\affiliation[CASIOP]
{Beijing National Laboratory for Condensed Matter Physics,
Institute of Physics, Chinese Academy of Sciences; School of Physical Sciences,
University of Chinese Academy of Sciences, Beijing 100190,China}

\author{Mo Chen}
\affiliation[BAQIS]
{Beijing Academy of Quantum Information Sciences, 100193 Beijing, China}

\author{Xingjun Wu}
\affiliation[BAQIS]
{Beijing Academy of Quantum Information Sciences, 100193 Beijing, China}

\author{Li Lu}
\affiliation[CASIOP]
{Beijing National Laboratory for Condensed Matter Physics,
Institute of Physics, Chinese Academy of Sciences; School of Physical Sciences,
University of Chinese Academy of Sciences, Beijing 100190,China}

\author{Jianhua Zhao}
\email{jhzhao@red.semi.ac.cn}
\affiliation[CASIOS]
{State Key Laboratory of Superlattices and Microstructures, Institute of Semiconductors,Chinese Academy of Sciences, P.O. Box 912, Beijing 100083, China}

\author{Ji-Yin Wang}
\email{wangjiyinshu@gmail.com}
\affiliation[BAQIS]
{Beijing Academy of Quantum Information Sciences, 100193 Beijing, China}

\author{H. Q. Xu}
\email{hqxu@pku.edu.cn}
\affiliation[PKU]
{Beijing Key Laboratory of Quantum Devices, Key Laboratory for the Physics and Chemistry of Nanodevices, and School of Electronics, Peking University, Beijing 100871, China}
\alsoaffiliation[BAQIS]
{Beijing Academy of Quantum Information Sciences, 100193 Beijing, China}

\title{Supercurrent, Multiple Andreev Reflections and Shapiro Steps in InAs Nanosheet Josephson Junctions }


\begin{document}
\newpage







\begin{abstract}
High-quality free-standing InAs nanosheets are emerging layered semiconductor materials with potentials in  designing planar Josephson junction devices for novel physics studies due to their unique properties including strong spin-orbit couplings, large Landé g-factors and the two dimensional nature. Here, we report an experimental study of proximity induced superconductivity in planar Josephson junction devices made from free-standing InAs nanosheets. The nanosheets are grown by molecular beam epitaxy and the Josephson junction devices are fabricated by directly contacting the nanosheets with superconductor Al electrodes. The fabricated devices are explored by low-temperature carrier transport measurements. The measurements show that the devices exhibit a gate-tunable supercurrent, multiple Andreev reflections, and a good quality superconductor-semiconductor interface. The superconducting characteristics of the Josephson junctions are investigated at different magnetic fields and temperatures, and are analyzed based on the Bardeen-Cooper-Schrieffer (BCS) theory. The measurements of ac Josephson effect are also conducted under microwave radiations with different radiation powers and frequencies, and integer Shapiro steps are observed. Our work demonstrates that InAs nanosheet based hybrid devices are desired systems for investigating forefront physics, such as the two-dimensional topological superconductivity.
\end{abstract}
\newpage

\section{Introduction}
\par Hybrid superconducting devices made from low-dimensional semiconductors with strong spin-orbit couplings, such as InAs and InSb, have attracted great research interests. These devices have now been widely explored for the investigation of exotic physics including anomalous Josephson effect, \cite{yokoyama2014anomalous,szombati2016josephson,strambini2020josephson,mayer2020gate} Josephson diode effect,\cite{chen2018asymmetric,baumgartner2022supercurrent,turini2022josephson,mazur2022gate} and topological superconductivity,  \cite{lutchyn2010majorana,oreg2010helical,mourik2012signatures,deng2012anomalous,deng2014,pientka2017topological,fornieri2019evidence,ren2019topological,aghaee2022inas} and for their potential applications in topological quantum computation. \cite{nayak2008non,stern2013topological,schrade2018majorana} Many excellent progresses have been achieved in devices made from InAs or InSb nanowires.\cite{Nilsson2009,Nilsson2010,mourik2012signatures,nilsson2012supercurrent,abay2012high,deng2012anomalous,abay2013,szombati2016josephson,albrecht2016exponential}. Comparing to nanowire-based hybrid devices, two-dimensional planar Josephson junctions have the advantage of flexibility for scaling up and it has been shown that the topological phase can be tuned by the phase difference across a planar Josephson junction. \cite{pientka2017topological,fornieri2019evidence,ren2019topological} In addition to nanowires and two-dimensional electron gases (2DEGs) formed in heterostructures, free-standing nanosheets have shown up as an emerging alternative material platform \cite{pan2016free,de2016twin,pan2019dimension,sun2020high}. More than having the same superiorities as 2DEGs mentioned above, nanosheets are capable of being transferred freely, facilitating the fabrication of dual-gate devices. Importantly, spin-orbit interaction, a key parameter in forefront areas of physics, can be tuned by a vertical electric field applied through InAs and InSb nanosheets with the dual-gate architecture. \cite{chen2021strong,fan2022electrically}. One of the prerequisite conditions to study the diverse physics in semiconductor-superconductor hybrid systems is the induced superconductivity in the semiconductors by proximity effect. Recently, proximity induced superconductivity has been realized in InSb nanosheet Josephson junctions\cite{kang2018two,zhi2019supercurrent,salimian2021gate,turini2022josephson,iorio2023half}.  Free-standing InAs nanosheets, on the other hand, appear as contestant materials with the similar intriguing properties as InSb nanosheets.\cite{pan2019dimension,fan2020measurements,fan2022electrically} However, the proximity induced superconductivity in a free-standing InAs nanosheet, which deserves to be demonstrated as an initial step, has not yet been reported.

\par Here, we report an experimental study of Josephson junctions made from high-quality, free-standing InAs nanosheets. These nanosheets are grown by molecular-beam epitaxy (MBE) and the devices are fabricated by transferring the nanosheets to a Si/SiO$_2$ substrate and then by directly contacting the nanosheets with superconductor Al electrodes. The fabricated InAs nanosheet Josephson junction devices are studied by low-temperature carrier transport measurements. The measurements show that these Josephson junction devices exhibit a gate tunable supercurrent (up to $\sim 50\,\mathrm{nA}$) and a good interface transparency. Multiple Andreev reflections (MARs) with clear subharmonic peaks at $n=\pm1,\, \pm2,\, \pm4$ have shown up. The magnetic field and temperature dependencies of the induced superconductivity are studied, from which a superconducting energy gap of $\Delta\sim 150\,\mathrm{\mu eV}$, a critical temperature of $T_\mathrm{c}\sim 1.05\,\mathrm{K}$ and a critical magnetic field of $B_\mathrm{c}\sim 9.6\,\mathrm{mT}$ are extracted. The InAs nanosheet Josephson junction devices are also investigated under microwave radiations and the ac Josephson effects with clear integer Shapiro steps (up to $n=\pm\,5$) are observed. These experimental results show that InAs nanosheets are an excellent material platform for investigating novel superconducting physics, including topological superconductivity.

\section{Results and discussion}

\par
\begin{figure*}[!h] 
\centering
\includegraphics[width=1\linewidth] {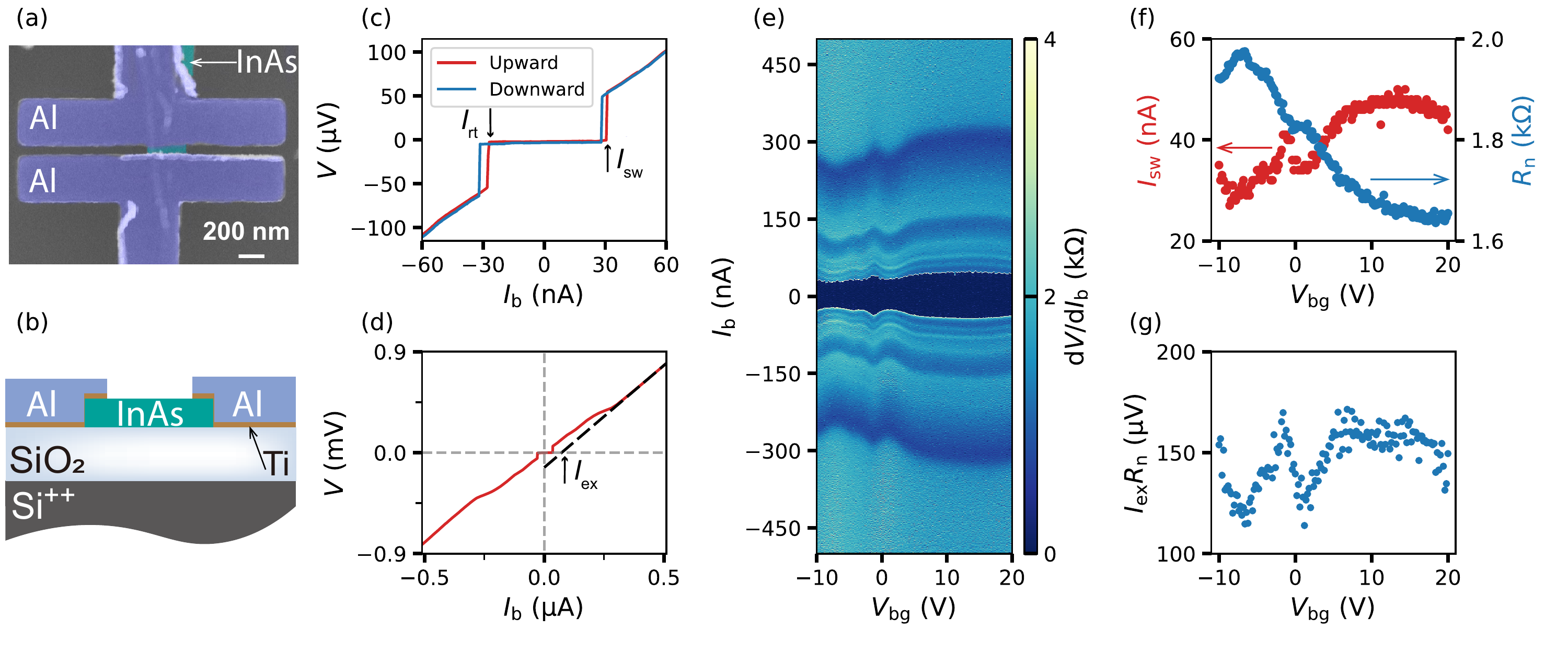}
\caption{\doublespacing \textbf{(a)} False colored SEM image of an InAs nanosheet Josephson junction device with the measurement data presented in the main article. The device is made on a heavily n-doped Si/SiO$_2$ substrate. The InAs nanosheet is colored in green. The source and drain electrodes are colored in light blue and are made of Ti/Al ($5\,\mathrm{nm}/80\,\mathrm{nm}$ in thickness). The scale bar is $200\,\mathrm{nm}$. \textbf{(b)} Cross-sectional schematic view of the device. \textbf{(c)} Measured voltage $V$ across the junction as a function of bias current $I_{\mathrm{b}}$ at back gate voltage $V_{\mathrm{bg}}=0$. The red and blue lines represent the measurements taken with the upward and downward current sweeping directions, respectively. The switching current $I_{\mathrm{sw}}$ and the retrapping current $I_{\mathrm{re}}$ are indicated with black arrows. \textbf{(d)} $V$-$I_{\mathrm{b}}$ curve in a larger $I_{\mathrm{b}}$ sweeping range. The black dashed line is a linear fit the $V$-$I_{\mathrm{b}}$ curve in a range where $V$ is larger than $2{\Delta}/e$. The fitting line extrapolates to a finite current (i.e., excess current $I_{\mathrm{ex}}$) at $V=0$, while the slope of the fitting line gives the resistance $R_\mathrm{n}$ of the junction in the normal state. \textbf{(e)} Differential resistance $dV$/$dI_{\mathrm{b}}$, on a color scale, as a function of $V_{\mathrm{bg}}$ and $I_{\mathrm{b}}$ (upward current sweeping direction only). The central dark area is the region of $dV$/$I_{\mathrm{b}} =0$ where the Josephson junction is in the superconducting state. \textbf{(f)}  Switching current  $I_{\mathrm{sw}}$ (red dots) and normal state resistance $R_{\mathrm{n}}$ (blue dots) extracted from the measurements in (e) as a function of $V_{\mathrm{bg}}$. \textbf{(g)} $I_{\mathrm{ex}}R_{\mathrm{n}}$ product extracted from (e) as a function of $V_{\mathrm{bg}}$. All data here is measured at $B=0$ and base temperature $T\sim 20\,\mathrm{mK}$.} \label{figure:1}
\end{figure*}

 The exploited free-standing InAs nanosheets grown by MBE are in wurtzite crystal structure with a thickness ranging from $15$ to $30\,\mathrm{nm}$ (see Ref. 29 for the details of the InAs nanosheet growth).  Figure \ref{figure:1}a shows a false-colored scanning electron microscope (SEM) image of a device measured in this work, where the scale bar is $200\,\mathrm{nm}$. Figure \ref{figure:1}b shows a cross-sectional schematic view of the device. The InAs nanosheets are mechanically transferred onto a highly n-doped Si substrate, covered by a 300-nm-thick $\mathrm{{SiO_\mathrm{2}}}$ layer on top, which serve as a global back gate in the following measurements. Standard electron-beam lithography is used to pattern the contact areas of the devices. In-situ argon ion milling has been conducted to remove the oxide layer on the surface of the InAs nanosheets prior to deposition of 5-nm-thick Ti and 80-nm-thick Al with electron-beam evaporation. The width of the nanosheet at the junction of a device is $\sim 300\,\mathrm{nm}$ and the gap between the two superconductor electrode is $\sim 90\,\mathrm{nm}$. Electrical measurements on the devices have been performed in a $^3\mathrm{He/}^4\mathrm{He}$ dilution refrigerator at a base temperature of $\sim 20\,\mathrm{mK}$. Two comparable devices have been explored in this work. The results shown in the main article all come from the InAs nanosheet Josephson junction device shown in Figure \ref{figure:1}a and the data measured for the other device are presented in the Supporting Information.

\par Figure \ref{figure:1}c shows the measured voltage $V$ across the junction of the hybrid device as a function of bias current $I_{\mathrm{b}}$ in a quasi-four-terminal measurement circuit setup (see details of the measurement setup in Figure S2 in the Supporting Information). The red and blue curves present the results of the measurements in the upward and downward bias current  sweeping directions, respectively. Both $V-I_{\mathrm{b}}$ curves show a clear zero voltage region due to a dissipationless supercurrent flow. The switching current $I_\mathrm{sw}$ and the retrapping current $I_\mathrm{rt}$, which are indicated by black arrows, are found to be around $30\,\mathrm{nA}$ and  $26\,\mathrm{nA}$, respectively, in the device. A hysteretic behavior seen from the upward and downward sweeping measurements, which is commonly observed in nanostructure Josephson junctions \cite{doh2005tunable,nilsson2012supercurrent}, can be explained by phase instability in the junction and/or heating effects.  \cite{tinkham2004introduction,tinkham2003hysteretic,courtois2008origin} Figure \ref{figure:1}d displays a $V-I_{\mathrm{b}}$ trace measured in a larger $I_{\mathrm{b}}$ range (upward sweeping direction only). At sufficiently large $I_{\mathrm{b}}$, where a voltage drop through the junction is larger than $2\Delta/e$ with $\Delta$ being the superconducting gap and e the elementary charge, $V$ appears to show a linear dependence on $I_{\mathrm{b}}$. By a linear extrapolation from the linear range of the $V-I_{\mathrm{b}}$ curve, excess current $I_{\mathrm{ex}}$, which originates from Cooper pair transport, \cite{bratus1995theory,cuevas1996hamiltonian} can be obtained. At the same time, the normal state resistance $R_{\mathrm{n}}$ is extracted from the slope of the linear fit. Here, in Figure \ref{figure:1}d, it is found that $I_{\mathrm{ex}}$ is about $72\,\mathrm{nA}$ and $R_{\mathrm{n}}$ is around $1.82\,\mathrm{k\Omega}$. Figure \ref{figure:1}e shows the differential resistance ${d}V/{d}I_\mathrm{b}$ of the device as a function of $I_{\mathrm{b}}$ and back gate voltage $V_\mathrm{bg}$ (extracted from the upward current sweeping measurements). The center dark area represents that the junction is in the superconducting state and  the dark fringes caused by MARs are clearly visible at the outside of the center dark area. The device also exhibits a gate-tunable $I_\mathrm{sw}$, which ranges from $27$ to $50\,\mathrm{nA}$ in the varying region of $V_{\mathrm{bg}}$ from $-10$ to $20\,\mathrm{V}$. Figure \ref{figure:1}f displays $I_{sw}$ (red data point) and $R_{\mathrm{n}}$ (blue data point) extracted from Figure \ref{figure:1}e as a function $V_\mathrm{bg}$. The values of the product $I_{sw}R_{\mathrm{n}}$ obtained from the extracted values of $I_{sw}$ and $R_{\mathrm{n}}$ shown in Figure \ref{figure:1}f are in the range of $50$ to $80\,\mu V$. In a short, disordered Josephson junction, the product of $I_{c}R_{\mathrm{n}}$, where $I_{c}$ is the critical supercurrent, was predicted to be related to the superconducting gap $\Delta$ as $I_{c}R_{\mathrm{n}}= \pi\Delta/2e$\cite{beenakker1991universal,tinkham2004introduction}. The values of product $I_{sw}R_{\mathrm{n}}$ extracted from our InAs nanosheet Josephson junction device are significantly smaller than the value of $I_{c}R_{\mathrm{n}}\sim 230\,\mathrm{{\mu}V}$, which would be obtained by taking $\Delta\sim 150\,\mu eV$ (see Figure \ref{figure:3} below).  This discrepancy can be explained by a smaller $I_\mathrm{sw}$ as compared to $I_\mathrm{c}$ due to premature switching. \cite{doh2005tunable,nilsson2012supercurrent} 

\par Figure \ref{figure:1}g shows the value of the product $I_\mathrm{ex}R_{\mathrm{n}}$ as a function of $V_\mathrm{bg}$, where $I_\mathrm{ex}$ and $R_{\mathrm{n}}$ are extracted from the measurements shown in Figure \ref{figure:1}e (see Figure \ref{figure:1}d and the related text for the extractions of $I_\mathrm{ex}$ and $R_{\mathrm{n}}$). It is seen that the value of $I_\mathrm{ex}R_{\mathrm{n}}$ varies from 113 to 172 ${\mu}$V, which are more comparable to the value of $(\pi^{2}/4-1)\Delta/e\sim 220$ $\mu$V predicted for a short disordered junction\cite{artemenko1979theory} than the value of $8\Delta/3e\sim 400$ $\mu$V predicted for a short ballistic junction \cite{klapwijk1982explanation,flensberg1988subharmonic}. 
These experimentally extracted values of $I_\mathrm{ex}R_{\mathrm{n}}$ can be used to evaluate the transparency of our InAs nanosheet Josephson junctions. The transmission $T_{\mathrm{r}}$ of a Josephson junction can be estimated based on the formula of $T_{\mathrm{r}}=1/(1+Z^2)$, where $Z$ is the scattering parameter at the interface between the superconductor and the normal conductor and can be determined from the ratio of $eI_\mathrm{ex}R_\mathrm{n}/{\Delta}$.\cite{blonder1982transition,flensberg1988subharmonic}  In our device, this ratio is found to be in the range of $\sim 0.65$ to $\sim 0.5$, leading to the transmission  $T_{\mathrm{r}}$ in a range of $\sim 70\,{\%}$ to $\sim 80\,{\%}$. Similar values of $T_{\mathrm{r}}$ have been obtained in the second device (see Figure S4 in the Supporting Information). The transparencies of our InAs nanosheet Josephson junction devices are comparable with the values obtained in previous works\cite{doh2005tunable,li2016coherent,nishio2011supercurrent}.

\par
\begin{figure*}[!h] 
\centering
\includegraphics[width=0.5\linewidth] {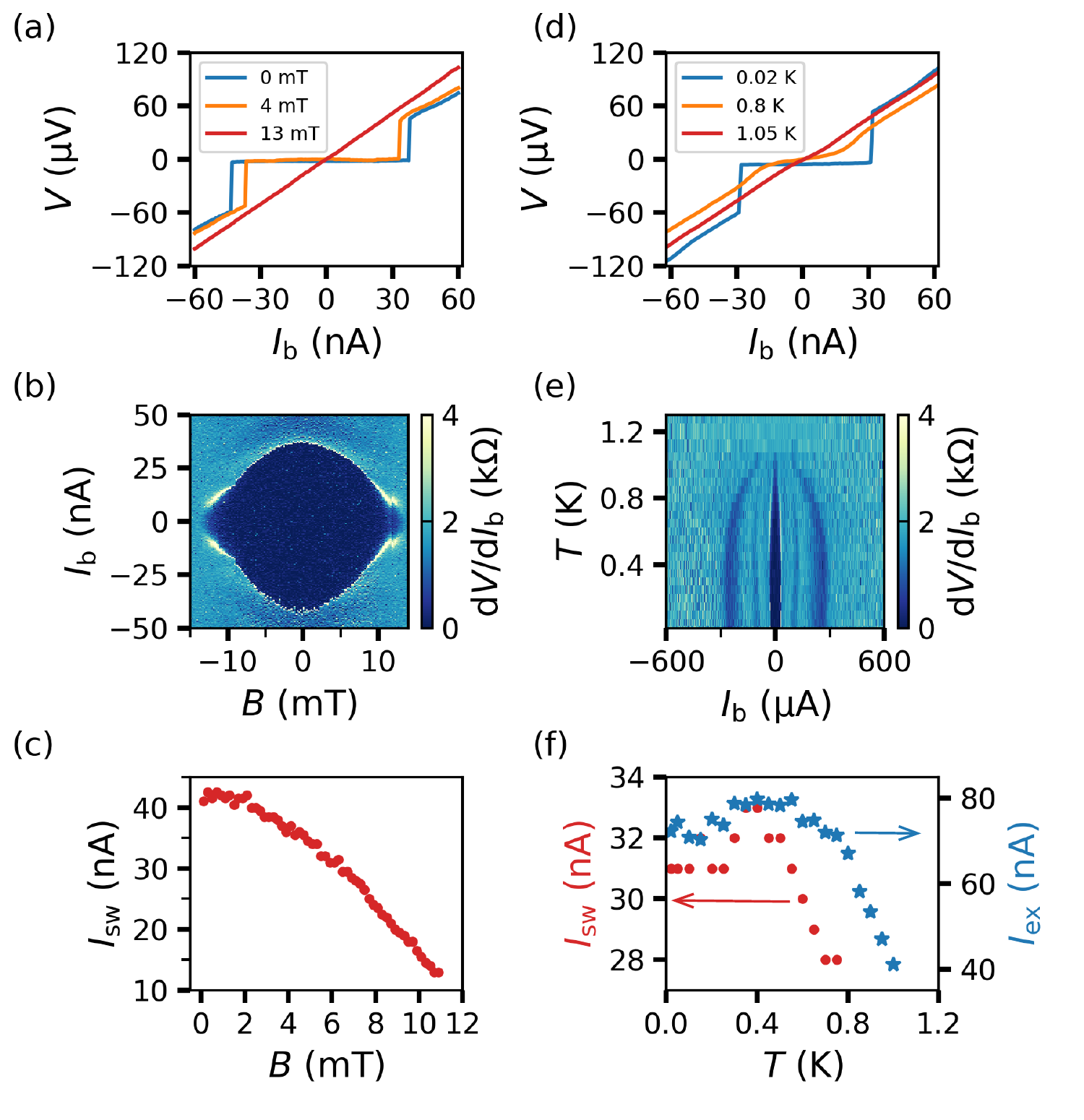}
\caption{\doublespacing \textbf{(a)} Measured voltage $V$ as a function of bias current $I_{\mathrm{b}}$ at different magnetic fields $B$. Here, at each magnetic field, only the measurements with sweeping current downward are shown. \textbf{(b)} Differential resistance ${dV}/dI_{\mathrm{b}}$, on a colour scale, as a function of magnetic field $B$ and bias current $I_{\mathrm{b}}$ (only the measurements with sweeping current downward are shown). The central dark blue area is the superconducting region with $dV/dI_{\mathrm{b}}=0$. The measurements data shown in (a) and (b) are taken at $V_{\mathrm{bg}}=20\,\mathrm{V}$ and base temperature $T\sim 20\,\mathrm{mK}$. \textbf{(c)} Switching current $I_{\mathrm{sw}}$ extracted from the measurement in (b) as a function of magnetic field $B$. \textbf{(d)} Measured voltage $V$ as a function of bias current $I_{\mathrm{b}}$ at different temperatures $T$. Here, at each temperature, only the measurements with sweeping current downward are shown. \textbf{(e)} Differential resistance $dV/dI_{\mathrm{b}}$, on a colour scale, as a function of bias current $I_{\mathrm{b}}$ (data with upward current sweeping direction only) and temperature $T$. The central dark blue area is the superconducting region with $dV/dI_{\mathrm{b}}=0$. The measurements data shown in (d) and (e) are taken at $V_{\mathrm{bg}}=0$ and $B=0$. \textbf{(f)} Switching current $I_{\mathrm{sw}}$ and excess current $I_{\mathrm{ex}}$ as a function of temperature $T$. $I_{\mathrm{sw}}$ is extracted from right the edge of the center dark area in (e) and $I_{\mathrm{ex}}$ is extracted from the measurements in (e) with the method illustrated in Figure 1d.} \label{figure:2}
\end{figure*}

 In Figure \ref{figure:2}, we examine the superconducting characteristics of the InAs nanosheet Josephson junctions at different magnetic fields and temperatures. Figure \ref{figure:2}a displays the measured voltage drop ${V}$ across the Josephson junction shown in Figure \ref{figure:1}a as a function of applied bias current $I_{\mathrm{b}}$ at magnetic fields $B=0$ (blue), $B=4\,\mathrm{mT}$ (orange) and $B=13\,\mathrm{mT}$ (red), applied perpendicularly to the InAs nanosheet, and at $V_{\mathrm{bg}}=20\,\mathrm{V}$. By comparing the three measured ${V}-I_{\mathrm{b}}$ curves, we see that the switching current  is decreased with increasing magnetic field. Figure \ref{figure:2}b shows the differential resistance $d{V}/d{I_{\mathrm{b}}}$ as a function of $I_{b}$ (downward sweeping direction only) and $B$ for the device at $V_{\mathrm{bg}}=20\,\mathrm{V}$. The center dark blue area presents the superconductive region, with the switching current $I_\mathrm{sw}$ given by the lower edge of the center area. Figure \ref{figure:2}c shows the switching current $I_\mathrm{sw}$ as a function of $B$. As shown in Figure \ref{figure:2}b and Figure \ref{figure:2}c, $I_\mathrm{sw}$ drops monotonically with increasing $B$, showing no oscillations. The failure of observing interference patterns could be due to a small junction area so that superconductivity is completely destroyed before the side lobes of a Fraunhofer pattern commence. Similar results are observed in the second device (see details in Figure S5 in the Supporting Information). Figure \ref{figure:2}d shows the measured voltage drop $V$ across the junction  in the device shown in Figure \ref{figure:1}a at three different temperatures $T=0.02\,\mathrm{K}$ (blue), $T=0.8\,\mathrm{K}$ (orange) and $T=1.05\,\mathrm{K}$ (red). As expected, $I_\mathrm{sw}$ is seen to be smaller at higher temperatures. Figure \ref{figure:2}e provides the differential resistance $dV/dI_\mathrm{b}$ as a function of $I_\mathrm{b}$ (upward sweeping direction only) and $T$ over a larger range of $I_\mathrm{b}$ and $T$. The center dark low-resistance area seen around $I_\mathrm{b}\sim 0$ represents that the junction is at the superconducting state and the other low resistance stripes seen at finite $I_\mathrm{b}$ originate from multiple Andreev reflections (MARs). Figure \ref{figure:2}f displays the switching current $I_\mathrm{sw}$ (red dots) and the excess current $I_\mathrm{ex}$ (blue dots) extracted from Figure \ref{figure:2}e as a function of temperature $T$ (cf.~Figure \ref{figure:1}d for the extraction of $I_\mathrm{ex}$). As expected, it is seen that both $I_\mathrm{sw}$ and $I_\mathrm{ex}$ decrease with increasing $T$ at high temperatures of $T > 0.4\,\mathrm{K}$. However, in the small temperature region of $T< 0.4\,\mathrm{K}$, both $I_\mathrm{sw}$ and $I_\mathrm{ex}$ are seen to raise with increasing $T$.  The origin of such anomalous enhancements of $I_\mathrm{sw}$ and $I_\mathrm{ex}$ is not known to us and thus deserves an investigation in the future.

\par We now turn to investigate the superconductivity in our InAs nanosheet Josephson junctions in a voltage bias measurement configuration. Figure \ref{figure:3}a shows the differential conductance $dI_{\mathrm{sd}}$/$dV_{\mathrm{sd}}$ measured for the device shown in Figure \ref{figure:1}a as a function of bias voltage $V_{\mathrm{sd}}$ at $T\sim\,20\,\mathrm{mK}$, $V_{\mathrm{bg}}=0\,\mathrm{V}$ and $B=0\,\mathrm{T}$. The data displays clear peaks with peak positions appearing approximately at $eV_{\mathrm{sd}}(n)=2\Delta/n$, where $n=\pm 1,\,\pm 2\, \mathrm{and}\, \pm 4$, indicated by the grey dashed lines in the figure. These peaks result from the MAR processes. The clear observation of MARs up to $n=\pm 4$ indicates a high quality and a high interface transparency in the InAs nanosheet Josephson junction. Figure \ref{figure:3}b shows a linear fit of peak position ($V_{\mathrm{peak}}$) versus the inversion of the MAR order ($1/n$). The superconducting energy gap extracted from these observed MAR peaks is ${\Delta}\sim 150\,\mathrm{\mu eV}$. Figure \ref{figure:3}c shows the differential conductance $dI_{\mathrm{sd}}$/$dV_{\mathrm{sd}}$ measured for the InAs nanosheet Josephson junction as a function of the bias voltage $V_{\mathrm{sd}}$ and temperature ${T}$. As $T$ increases, the peak positions move consecutively to lower values of $V_{\mathrm{sd}}$, due to the decrease in the superconducting energy gap $\Delta$ with increasing $T$. According to the prediction of the Bardeen-Cooper-Schrieffer (BCS) theory, the temperature dependence of the superconducting energy gap follows $\Delta(T)={\Delta(0)}[{cos} [{\pi/2}(T/T_\mathrm{c})^2]^{1/2}$, where $T_\mathrm{c}$ is the critical temperature of the superconductor.\cite{tinkham2004introduction} The black dashed lines are the fitting results for the positions of the first order peaks, $V_{\mathrm{sd}}=\pm 2\Delta/e$, based on the BCS theoretical prediction. The fitting gives a zero temperature superconducting energy gap of ${\Delta(0)}\sim 151\,\mathrm{\mu eV}$ and a critical temperature of  $T_\mathrm{c}\sim 1.05\,\mathrm{K}$. 
Figure \ref{figure:3}d shows the differential conductance $dI_{\mathrm{sd}}$/$dV_{\mathrm{sd}}$ measured for the InAs nanosheet Josephson junction as a function of bias voltage $V_{\mathrm{sd}}$ and magnetic field ${B}$. Again, the measurements show a few well-resolved MAR peak structures and their magnetic field  dependencies follow the evolution of the superconducting energy gap $\Delta$ predicted by the BCS theory, $\Delta(B)={\Delta(0)}[1-(B/B_c)^2]^{1/2}$, see the black dashed lines for the theoretical fits to the two first order MAR peaks. The extracted  zero field superconducting energy gap $\Delta(0)$ and the critical field ${B_{\mathrm{c}}}$ from the fits are ${\Delta(0)}\sim 154\,\mathrm{\mu eV}$ and ${B_{\mathrm{c}}}\sim 9.6\,\mathrm{mT}$. 

 \par
\begin{figure*}[!h] 
\centering
\includegraphics[width=\linewidth] {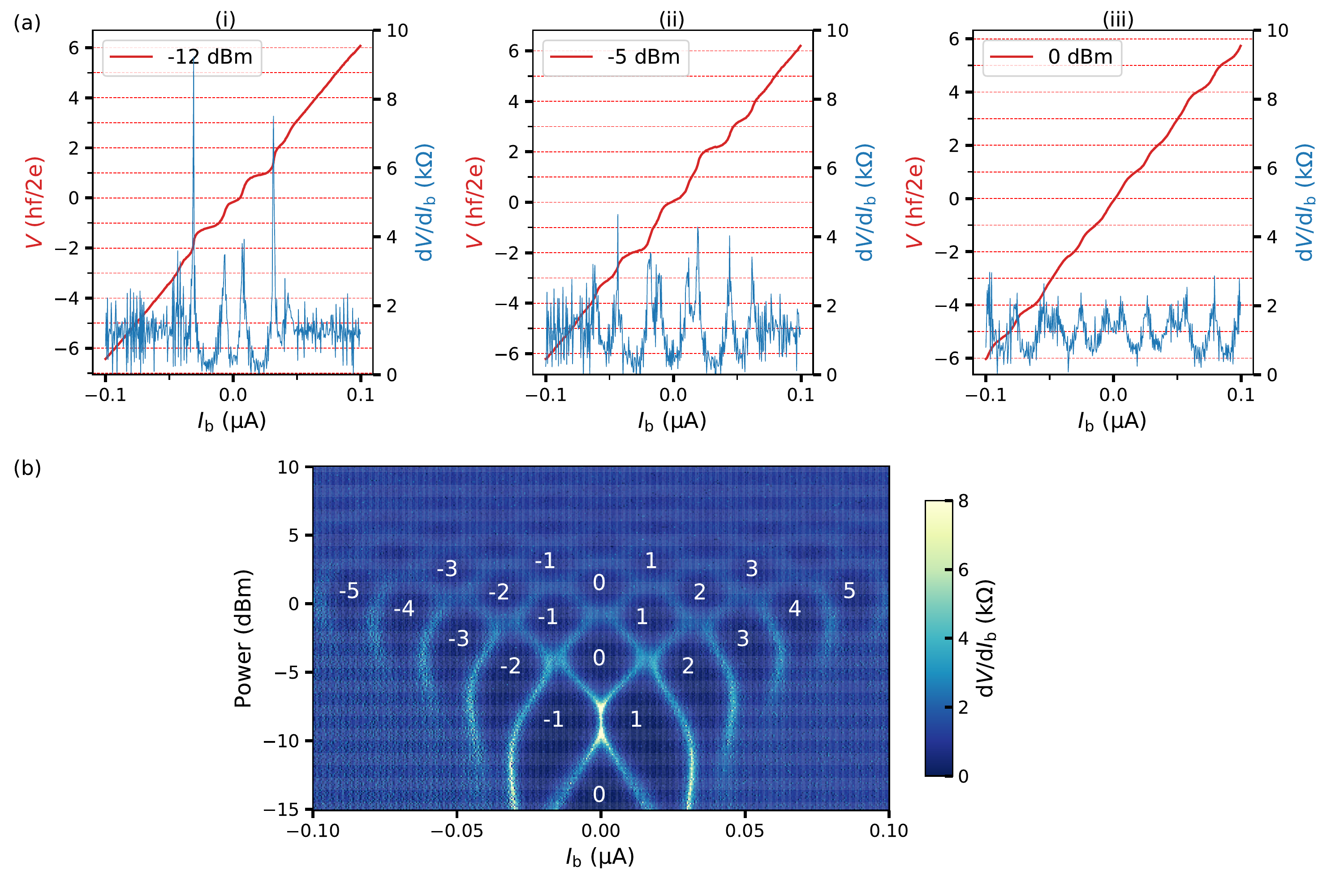}
\caption{\doublespacing ac Josephson effect measured under microwave radiation of frequency ${f}=10\,\mathrm{GHz}$ at $B=0$ and $T\sim\,20\,\mathrm{mK}$. \textbf{(a)} Measured voltage $V$ in units of $h\!f/2e$ (red lines) and extracted differential resistance $dV/dI_{\mathrm{b}}$ (blue lines) as a function of bias current $I_{\mathrm{b}}$ under microwave powers of (i) $-12\,\mathrm{dBm}$, (ii)  $-5\,\mathrm{dBm}$ and (iii) $0\,\mathrm{dBm}$. Here, the integer Shapiro steps of several orders n at voltages $V=nhf/2e$ are clearly visible. \textbf{(b)} Differential resistance $dV/dI_{\mathrm{b}}$, on a color scale, as a function of bias current $I_{\mathrm{b}}$ and microwave power. The dark blue areas marked with integer numbers $n$ are the regions where the integer Shapiro steps are found.}\label{figure:4}
\end{figure*}

 We have also examined our InAs nanosheet Josephson junction device under microwave radiation and studied the evolution of Shapiro steps at different radiation powers and frequencies. Shapiro step measurements have been used to examine the current-phase relation (CPR), where integer Shapiro steps represent a sinusoidal CPR with a ${2\pi}$ periodicity and half-integer or fractional indexed Shapiro steps indicate a skewed CPR. \cite{dinsmore2008fractional,raes2020fractional} Furthermore, the missing of odd-integer multiples of Shapiro steps implies a ${4\pi}$ periodical CPR, which could help to identify the presence of a topological superconducting phase in a Josephson junction.\cite{rokhinson2012fractional,wiedenmann20164,rosenbach2021reappearance,wiedenmann20164} Figure \ref{figure:4}a shows the responses of the InAs nanosheet Josephson junction device shown in Figure \ref{figure:1}a in the current bias setup to applied microwave radiations at a frequency of $f=10\,\mathrm{GHz}$ and powers of (i) $P=-12$ dBm, (ii) $P=-5$ dBm and (iii) $P=0$ dBm. The red lines are the measured voltage $V$ in units of $h\!f/2e$ as a function of applied bias current $I_\mathrm{b}$ and the blue lines are the extracted differential resistance $dV$/$dI_{\mathrm{b}}$. Clear integer voltage steps (integer Shapiro steps) appear at voltage values of ${nh\!f/2e}$, where $n=0,\,\pm1$ under the microwave power of $-12\,\mathrm{dBm}$, $n=0,\,\pm1,\,\pm2,\,\pm3$ under the microwave power of $-5\,\mathrm{dBm}$, and $n=0,\,\pm1,\,\pm2,\,\pm3\,\pm4,\,\pm5$ under the microwave power of $0\,\mathrm{dBm}$. Figure \ref{figure:4}b shows the differential resistance $dV$/$dI_{\mathrm{b}}$ extracted from the measurements of the device as a function of $I_{\mathrm{b}}$ and microwave power at $f=10\,\mathrm{GHz}$. The dark blue, low differential resistance areas marked with integer numbers $n$ are the regions where the integer Shapiro steps are observed. Here, the Shapiro steps up to $n=\pm\,5$ can be clearly seen. Such ac Josephson effect measurements have also been carried out at other frequencies for this device and for  the second device. Shapiro steps up to high orders can be identified as well in these measurements, see details in Figure S7 in the Supporting Information. The clearly observed integer Shapiro steps reveal the presence of a sinusoidal CPR with a ${2\pi}$ periodicity in our InAs nanosheet Josephson junction devices.
 
\par In summary, we have fabricated Al-InAs nanosheet-Al Josephson junction devices and investigated the proximity induced superconductivity in the devices. The switching current $I_\mathrm{sw}$ is tunable using a back gate and  a maximum value of $I_\mathrm{sw}\sim 50\,\mathrm{nA}$ is observed. The product of the excess current $I_\mathrm{ex}$ and the normal state resistance $R_\mathrm{n}$ as high as $\sim 170\,\mathrm{\mu V}$ is found. We have also observed the well resolved MAR structures of order $n=\pm1,\,\pm2,\,\pm4$, verifying a high quality and a high transparency of our Josephson junction devices. The ac Josephson effect is investigated in the devices under microwave radiation and integer Shapiro steps of order $n$ up to $\pm\,5$ are observed. The excellent proximity-induced superconducting properties observed in our InAs nanosheet Josephson junction devices show that free-standing InAs nanosheets can serve as a new material platform for hybrid devices designated for study of novel Josephson junction physics and for realization and manipulation of topological superconducting quantum states.

\section{Author contributions}

H.Q.X conceived and supervised the project. S.Y., H.S. and M.C. fabricated the devices with great aids from Z.L and L.L. D.P. and J.Z. grew the semiconductor InAs nanosheets. J.Y.W., W.L. and X.W. set up the microwave radiation measurement circuit in the dilution refrigerator. S.Y., J.Y.W. and H.S. performed the transport measurements. S.Y., J.Y.W. and H.Q.X. analyzed the measurement data. S.Y., J.Y.W. and H.Q.X. wrote the manuscript with inputs from all the authors.  

\begin{acknowledgement}

This work was supported by the National Natural Science Foundation of China (Grant Nos. 92165208, 11874071, 92065106, 61974138, and 12004039), the Ministry of Science and Technology of China through the National Key Research and Development Program of China (Grant Nos. 2017YFA0303304 and 2016YFA0300601). D.P. also acknowledges the supports from the Youth Innovation Promotion Association, Chinese Academy of Sciences (Nos. 2017156 and Y2021043).
\end{acknowledgement}

\section{Data analysis and data availability} 

All details in data analysis are included in a jupyter notebook file. The raw data and the analysis files are available at \url{https://doi.org/10.5281/zenodo.7837262}. 
      
\section{Conflict of interests}
The authors declare no conflict of interests. 

\bibliography{references}

\providecommand{\latin}[1]{#1}
\makeatletter
\providecommand{\doi}
  {\begingroup\let\do\@makeother\dospecials
  \catcode`\{=1 \catcode`\}=2 \doi@aux}
\providecommand{\doi@aux}[1]{\endgroup\texttt{#1}}
\makeatother
\providecommand*\mcitethebibliography{\thebibliography}
\csname @ifundefined\endcsname{endmcitethebibliography}
  {\let\endmcitethebibliography\endthebibliography}{}
\begin{mcitethebibliography}{56}
\providecommand*\natexlab[1]{#1}
\providecommand*\mciteSetBstSublistMode[1]{}
\providecommand*\mciteSetBstMaxWidthForm[2]{}
\providecommand*\mciteBstWouldAddEndPuncttrue
  {\def\EndOfBibitem{\unskip.}}
\providecommand*\mciteBstWouldAddEndPunctfalse
  {\let\EndOfBibitem\relax}
\providecommand*\mciteSetBstMidEndSepPunct[3]{}
\providecommand*\mciteSetBstSublistLabelBeginEnd[3]{}
\providecommand*\EndOfBibitem{}
\mciteSetBstSublistMode{f}
\mciteSetBstMaxWidthForm{subitem}{(\alph{mcitesubitemcount})}
\mciteSetBstSublistLabelBeginEnd
  {\mcitemaxwidthsubitemform\space}
  {\relax}
  {\relax}

\bibitem[Yokoyama \latin{et~al.}(2014)Yokoyama, Eto, and
  Nazarov]{yokoyama2014anomalous}
Yokoyama,~T.; Eto,~M.; Nazarov,~Y.~V. Anomalous Josephson effect induced by
  spin-orbit interaction and Zeeman effect in semiconductor nanowires.
  \emph{Phys. Rev. B} \textbf{2014}, \emph{89}, 195407\relax
\mciteBstWouldAddEndPuncttrue
\mciteSetBstMidEndSepPunct{\mcitedefaultmidpunct}
{\mcitedefaultendpunct}{\mcitedefaultseppunct}\relax
\EndOfBibitem
\bibitem[Szombati \latin{et~al.}(2016)Szombati, Nadj-Perge, Car, Plissard,
  Bakkers, and Kouwenhoven]{szombati2016josephson}
Szombati,~D.; Nadj-Perge,~S.; Car,~D.; Plissard,~S.; Bakkers,~E.;
  Kouwenhoven,~L. Josephson $\phi$ 0-junction in nanowire quantum dots.
  \emph{Nat. Phys.} \textbf{2016}, \emph{12}, 568--572\relax
\mciteBstWouldAddEndPuncttrue
\mciteSetBstMidEndSepPunct{\mcitedefaultmidpunct}
{\mcitedefaultendpunct}{\mcitedefaultseppunct}\relax
\EndOfBibitem
\bibitem[Strambini \latin{et~al.}(2020)Strambini, Iorio, Durante, Citro,
  Sanz-Fern{\'a}ndez, Guarcello, Tokatly, Braggio, Rocci, Ligato,
  \latin{et~al.} others]{strambini2020josephson}
Strambini,~E.; Iorio,~A.; Durante,~O.; Citro,~R.; Sanz-Fern{\'a}ndez,~C.;
  Guarcello,~C.; Tokatly,~I.~V.; Braggio,~A.; Rocci,~M.; Ligato,~N.,
  \latin{et~al.}  A Josephson phase battery. \emph{Nat. Nanotechnol.}
  \textbf{2020}, \emph{15}, 656--660\relax
\mciteBstWouldAddEndPuncttrue
\mciteSetBstMidEndSepPunct{\mcitedefaultmidpunct}
{\mcitedefaultendpunct}{\mcitedefaultseppunct}\relax
\EndOfBibitem
\bibitem[Mayer \latin{et~al.}(2020)Mayer, Dartiailh, Yuan, Wickramasinghe,
  Rossi, and Shabani]{mayer2020gate}
Mayer,~W.; Dartiailh,~M.~C.; Yuan,~J.; Wickramasinghe,~K.~S.; Rossi,~E.;
  Shabani,~J. Gate controlled anomalous phase shift in Al/InAs Josephson
  junctions. \emph{Nat. Commun.} \textbf{2020}, \emph{11}, 212\relax
\mciteBstWouldAddEndPuncttrue
\mciteSetBstMidEndSepPunct{\mcitedefaultmidpunct}
{\mcitedefaultendpunct}{\mcitedefaultseppunct}\relax
\EndOfBibitem
\bibitem[Chen \latin{et~al.}(2018)Chen, He, Ali, Lee, Fong, and
  Law]{chen2018asymmetric}
Chen,~C.-Z.; He,~J.~J.; Ali,~M.~N.; Lee,~G.-H.; Fong,~K.~C.; Law,~K.~T.
  Asymmetric Josephson effect in inversion symmetry breaking topological
  materials. \emph{Phys. Rev. B} \textbf{2018}, \emph{98}, 075430\relax
\mciteBstWouldAddEndPuncttrue
\mciteSetBstMidEndSepPunct{\mcitedefaultmidpunct}
{\mcitedefaultendpunct}{\mcitedefaultseppunct}\relax
\EndOfBibitem
\bibitem[Baumgartner \latin{et~al.}(2022)Baumgartner, Fuchs, Costa, Reinhardt,
  Gronin, Gardner, Lindemann, Manfra, Faria~Junior, Kochan, \latin{et~al.}
  others]{baumgartner2022supercurrent}
Baumgartner,~C.; Fuchs,~L.; Costa,~A.; Reinhardt,~S.; Gronin,~S.;
  Gardner,~G.~C.; Lindemann,~T.; Manfra,~M.~J.; Faria~Junior,~P.~E.;
  Kochan,~D., \latin{et~al.}  Supercurrent rectification and magnetochiral
  effects in symmetric Josephson junctions. \emph{Nat. Nanotechnol.}
  \textbf{2022}, \emph{17}, 39--44\relax
\mciteBstWouldAddEndPuncttrue
\mciteSetBstMidEndSepPunct{\mcitedefaultmidpunct}
{\mcitedefaultendpunct}{\mcitedefaultseppunct}\relax
\EndOfBibitem
\bibitem[Turini \latin{et~al.}(2022)Turini, Salimian, Carrega, Iorio,
  Strambini, Giazotto, Zannier, Sorba, and Heun]{turini2022josephson}
Turini,~B.; Salimian,~S.; Carrega,~M.; Iorio,~A.; Strambini,~E.; Giazotto,~F.;
  Zannier,~V.; Sorba,~L.; Heun,~S. Josephson Diode Effect in High-Mobility InSb
  Nanoflags. \emph{Nano Lett.} \textbf{2022}, \emph{22}, 8502--8508\relax
\mciteBstWouldAddEndPuncttrue
\mciteSetBstMidEndSepPunct{\mcitedefaultmidpunct}
{\mcitedefaultendpunct}{\mcitedefaultseppunct}\relax
\EndOfBibitem
\bibitem[Mazur \latin{et~al.}(2022)Mazur, van Loo, van Driel, Wang, Badawy,
  Gazibegovic, Bakkers, and Kouwenhoven]{mazur2022gate}
Mazur,~G.; van Loo,~N.; van Driel,~D.; Wang,~J.-Y.; Badawy,~G.;
  Gazibegovic,~S.; Bakkers,~E.; Kouwenhoven,~L. The gate-tunable Josephson
  diode. \emph{arXiv:2211.14283} \textbf{2022}, \relax
\mciteBstWouldAddEndPunctfalse
\mciteSetBstMidEndSepPunct{\mcitedefaultmidpunct}
{}{\mcitedefaultseppunct}\relax
\EndOfBibitem
\bibitem[Lutchyn \latin{et~al.}(2010)Lutchyn, Sau, and
  Sarma]{lutchyn2010majorana}
Lutchyn,~R.~M.; Sau,~J.~D.; Sarma,~S.~D. Majorana fermions and a topological
  phase transition in semiconductor-superconductor heterostructures.
  \emph{Phys. Rev. Lett.} \textbf{2010}, \emph{105}, 077001\relax
\mciteBstWouldAddEndPuncttrue
\mciteSetBstMidEndSepPunct{\mcitedefaultmidpunct}
{\mcitedefaultendpunct}{\mcitedefaultseppunct}\relax
\EndOfBibitem
\bibitem[Oreg \latin{et~al.}(2010)Oreg, Refael, and Von~Oppen]{oreg2010helical}
Oreg,~Y.; Refael,~G.; Von~Oppen,~F. Helical liquids and Majorana bound states
  in quantum wires. \emph{Phys. Rev. Lett.} \textbf{2010}, \emph{105},
  177002\relax
\mciteBstWouldAddEndPuncttrue
\mciteSetBstMidEndSepPunct{\mcitedefaultmidpunct}
{\mcitedefaultendpunct}{\mcitedefaultseppunct}\relax
\EndOfBibitem
\bibitem[Mourik \latin{et~al.}(2012)Mourik, Zuo, Frolov, Plissard, Bakkers, and
  Kouwenhoven]{mourik2012signatures}
Mourik,~V.; Zuo,~K.; Frolov,~S.~M.; Plissard,~S.; Bakkers,~E.~P.;
  Kouwenhoven,~L.~P. Signatures of Majorana fermions in hybrid
  superconductor-semiconductor nanowire devices. \emph{Science} \textbf{2012},
  \emph{336}, 1003--1007\relax
\mciteBstWouldAddEndPuncttrue
\mciteSetBstMidEndSepPunct{\mcitedefaultmidpunct}
{\mcitedefaultendpunct}{\mcitedefaultseppunct}\relax
\EndOfBibitem
\bibitem[Deng \latin{et~al.}(2012)Deng, Yu, Huang, Larsson, Caroff, and
  Xu]{deng2012anomalous}
Deng,~M.~T.; Yu,~C.~L.; Huang,~G.~Y.; Larsson,~M.; Caroff,~P.; Xu,~H.~Q.
  Anomalous zero-bias conductance peak in a Nb--InSb nanowire--Nb hybrid
  device. \emph{Nano Lett.} \textbf{2012}, \emph{12}, 6414--6419\relax
\mciteBstWouldAddEndPuncttrue
\mciteSetBstMidEndSepPunct{\mcitedefaultmidpunct}
{\mcitedefaultendpunct}{\mcitedefaultseppunct}\relax
\EndOfBibitem
\bibitem[Deng \latin{et~al.}(2014)Deng, Yu, Huang, Larsson, Caroff, and
  Xu]{deng2014}
Deng,~M.~T.; Yu,~C.~L.; Huang,~G.~Y.; Larsson,~M.; Caroff,~P.; Xu,~H.~Q. Parity
  independence of the zero-bias conductance peak in a nanowire based
  topological superconductor-quantum dot hybrid device. \emph{Sci. Rep.}
  \textbf{2014}, \emph{4}, 7261\relax
\mciteBstWouldAddEndPuncttrue
\mciteSetBstMidEndSepPunct{\mcitedefaultmidpunct}
{\mcitedefaultendpunct}{\mcitedefaultseppunct}\relax
\EndOfBibitem
\bibitem[Pientka \latin{et~al.}(2017)Pientka, Keselman, Berg, Yacoby, Stern,
  and Halperin]{pientka2017topological}
Pientka,~F.; Keselman,~A.; Berg,~E.; Yacoby,~A.; Stern,~A.; Halperin,~B.~I.
  Topological superconductivity in a planar Josephson junction. \emph{Phys.
  Rev. X} \textbf{2017}, \emph{7}, 021032\relax
\mciteBstWouldAddEndPuncttrue
\mciteSetBstMidEndSepPunct{\mcitedefaultmidpunct}
{\mcitedefaultendpunct}{\mcitedefaultseppunct}\relax
\EndOfBibitem
\bibitem[Fornieri \latin{et~al.}(2019)Fornieri, Whiticar, Setiawan,
  Portol{\'e}s, Drachmann, Keselman, Gronin, Thomas, Wang, Kallaher,
  \latin{et~al.} others]{fornieri2019evidence}
Fornieri,~A.; Whiticar,~A.~M.; Setiawan,~F.; Portol{\'e}s,~E.;
  Drachmann,~A.~C.; Keselman,~A.; Gronin,~S.; Thomas,~C.; Wang,~T.;
  Kallaher,~R., \latin{et~al.}  Evidence of topological superconductivity in
  planar Josephson junctions. \emph{Nature} \textbf{2019}, \emph{569},
  89--92\relax
\mciteBstWouldAddEndPuncttrue
\mciteSetBstMidEndSepPunct{\mcitedefaultmidpunct}
{\mcitedefaultendpunct}{\mcitedefaultseppunct}\relax
\EndOfBibitem
\bibitem[Ren \latin{et~al.}(2019)Ren, Pientka, Hart, Pierce, Kosowsky, Lunczer,
  Schlereth, Scharf, Hankiewicz, Molenkamp, \latin{et~al.}
  others]{ren2019topological}
Ren,~H.; Pientka,~F.; Hart,~S.; Pierce,~A.~T.; Kosowsky,~M.; Lunczer,~L.;
  Schlereth,~R.; Scharf,~B.; Hankiewicz,~E.~M.; Molenkamp,~L.~W.,
  \latin{et~al.}  Topological superconductivity in a phase-controlled Josephson
  junction. \emph{Nature} \textbf{2019}, \emph{569}, 93--98\relax
\mciteBstWouldAddEndPuncttrue
\mciteSetBstMidEndSepPunct{\mcitedefaultmidpunct}
{\mcitedefaultendpunct}{\mcitedefaultseppunct}\relax
\EndOfBibitem
\bibitem[Aghaee \latin{et~al.}(2022)Aghaee, Akkala, Alam, Ali, Ramirez,
  Andrzejczuk, Antipov, Astafev, Bauer, Becker, \latin{et~al.}
  others]{aghaee2022inas}
Aghaee,~M.; Akkala,~A.; Alam,~Z.; Ali,~R.; Ramirez,~A.~A.; Andrzejczuk,~M.;
  Antipov,~A.~E.; Astafev,~M.; Bauer,~B.; Becker,~J., \latin{et~al.}  InAs-Al
  hybrid devices passing the topological gap protocol. \emph{arXiv:2207.02472}
  \textbf{2022}, \relax
\mciteBstWouldAddEndPunctfalse
\mciteSetBstMidEndSepPunct{\mcitedefaultmidpunct}
{}{\mcitedefaultseppunct}\relax
\EndOfBibitem
\bibitem[Nayak \latin{et~al.}(2008)Nayak, Simon, Stern, Freedman, and
  Sarma]{nayak2008non}
Nayak,~C.; Simon,~S.~H.; Stern,~A.; Freedman,~M.; Sarma,~S.~D. Non-Abelian
  anyons and topological quantum computation. \emph{Rev. Mod. Phys.}
  \textbf{2008}, \emph{80}, 1083\relax
\mciteBstWouldAddEndPuncttrue
\mciteSetBstMidEndSepPunct{\mcitedefaultmidpunct}
{\mcitedefaultendpunct}{\mcitedefaultseppunct}\relax
\EndOfBibitem
\bibitem[Stern and Lindner(2013)Stern, and Lindner]{stern2013topological}
Stern,~A.; Lindner,~N.~H. Topological quantum computation-from basic concepts
  to first experiments. \emph{Science} \textbf{2013}, \emph{339},
  1179--1184\relax
\mciteBstWouldAddEndPuncttrue
\mciteSetBstMidEndSepPunct{\mcitedefaultmidpunct}
{\mcitedefaultendpunct}{\mcitedefaultseppunct}\relax
\EndOfBibitem
\bibitem[Schrade and Fu(2018)Schrade, and Fu]{schrade2018majorana}
Schrade,~C.; Fu,~L. Majorana superconducting qubit. \emph{Phys. Rev. Lett.}
  \textbf{2018}, \emph{121}, 267002\relax
\mciteBstWouldAddEndPuncttrue
\mciteSetBstMidEndSepPunct{\mcitedefaultmidpunct}
{\mcitedefaultendpunct}{\mcitedefaultseppunct}\relax
\EndOfBibitem
\bibitem[Nilsson \latin{et~al.}(2009)Nilsson, Caroff, Thelander, Larsson,
  Wagner, Wernersson, Samuelson, and Xu]{Nilsson2009}
Nilsson,~H.~A.; Caroff,~P.; Thelander,~C.; Larsson,~M.; Wagner,~J.~B.;
  Wernersson,~L.-E.; Samuelson,~L.; Xu,~H.~Q. Giant, Level-Dependent g Factors
  in InSb Nanowire Quantum Dots. \emph{Nano Lett.} \textbf{2009}, \emph{9},
  3151--3156\relax
\mciteBstWouldAddEndPuncttrue
\mciteSetBstMidEndSepPunct{\mcitedefaultmidpunct}
{\mcitedefaultendpunct}{\mcitedefaultseppunct}\relax
\EndOfBibitem
\bibitem[Nilsson \latin{et~al.}(2010)Nilsson, Caroff, Thelander, Larsson,
  Wagner, Wernersson, Samuelson, and Xu]{Nilsson2010}
Nilsson,~H.~A.; Caroff,~P.; Thelander,~C.; Larsson,~M.; Wagner,~J.~B.;
  Wernersson,~L.-E.; Samuelson,~L.; Xu,~H.~Q. Correlation-Induced Conductance
  Suppression at Level Degeneracy in a Quantum Dot. \emph{Phys. Rev. Lett.}
  \textbf{2010}, \emph{104}, 186804\relax
\mciteBstWouldAddEndPuncttrue
\mciteSetBstMidEndSepPunct{\mcitedefaultmidpunct}
{\mcitedefaultendpunct}{\mcitedefaultseppunct}\relax
\EndOfBibitem
\bibitem[Nilsson \latin{et~al.}(2012)Nilsson, Samuelsson, Caroff, and
  Xu]{nilsson2012supercurrent}
Nilsson,~H.~A.; Samuelsson,~P.; Caroff,~P.; Xu,~H.~Q. Supercurrent and multiple
  Andreev reflections in an InSb nanowire Josephson junction. \emph{Nano Lett.}
  \textbf{2012}, \emph{12}, 228--233\relax
\mciteBstWouldAddEndPuncttrue
\mciteSetBstMidEndSepPunct{\mcitedefaultmidpunct}
{\mcitedefaultendpunct}{\mcitedefaultseppunct}\relax
\EndOfBibitem
\bibitem[Abay \latin{et~al.}(2012)Abay, Nilsson, Wu, Xu, Wilson, and
  Delsing]{abay2012high}
Abay,~S.; Nilsson,~H.~A.; Wu,~F.; Xu,~H.~Q.; Wilson,~C.~M.; Delsing,~P. High
  critical-current superconductor-InAs nanowire-superconductor junctions.
  \emph{Nano Lett.} \textbf{2012}, \emph{12}, 5622--5625\relax
\mciteBstWouldAddEndPuncttrue
\mciteSetBstMidEndSepPunct{\mcitedefaultmidpunct}
{\mcitedefaultendpunct}{\mcitedefaultseppunct}\relax
\EndOfBibitem
\bibitem[Abay \latin{et~al.}(2013)Abay, Persson, Nilsson, Xu, Fogelstr\"{o}m,
  Shumeiko, and Delsing]{abay2013}
Abay,~S.; Persson,~D.; Nilsson,~H.; Xu,~H.~Q.; Fogelstr\"{o}m,~M.;
  Shumeiko,~V.; Delsing,~P. Quantized conductance and its correlation to the
  supercurrent in a nanowire connected to superconductors. \emph{Nano Lett.}
  \textbf{2013}, \emph{13}, 3614--3617\relax
\mciteBstWouldAddEndPuncttrue
\mciteSetBstMidEndSepPunct{\mcitedefaultmidpunct}
{\mcitedefaultendpunct}{\mcitedefaultseppunct}\relax
\EndOfBibitem
\bibitem[Albrecht \latin{et~al.}(2016)Albrecht, Higginbotham, Madsen, Kuemmeth,
  Jespersen, Nyg{\aa}rd, Krogstrup, and Marcus]{albrecht2016exponential}
Albrecht,~S.~M.; Higginbotham,~A.~P.; Madsen,~M.; Kuemmeth,~F.;
  Jespersen,~T.~S.; Nyg{\aa}rd,~J.; Krogstrup,~P.; Marcus,~C. Exponential
  protection of zero modes in Majorana islands. \emph{Nature} \textbf{2016},
  \emph{531}, 206--209\relax
\mciteBstWouldAddEndPuncttrue
\mciteSetBstMidEndSepPunct{\mcitedefaultmidpunct}
{\mcitedefaultendpunct}{\mcitedefaultseppunct}\relax
\EndOfBibitem
\bibitem[Pan \latin{et~al.}(2016)Pan, Fan, Kang, Zhi, Yu, Xu, and
  Zhao]{pan2016free}
Pan,~D.; Fan,~D.; Kang,~N.; Zhi,~J.; Yu,~X.; Xu,~H.~Q.; Zhao,~J. Free-standing
  two-dimensional single-crystalline InSb nanosheets. \emph{Nano Lett.}
  \textbf{2016}, \emph{16}, 834--841\relax
\mciteBstWouldAddEndPuncttrue
\mciteSetBstMidEndSepPunct{\mcitedefaultmidpunct}
{\mcitedefaultendpunct}{\mcitedefaultseppunct}\relax
\EndOfBibitem
\bibitem[De~La~Mata \latin{et~al.}(2016)De~La~Mata, Leturcq, Plissard, Rolland,
  Mag{\'e}n, Arbiol, and Caroff]{de2016twin}
De~La~Mata,~M.; Leturcq,~R.; Plissard,~S.~R.; Rolland,~C.; Mag{\'e}n,~C.;
  Arbiol,~J.; Caroff,~P. Twin-induced InSb nanosails: A convenient high
  mobility quantum system. \emph{Nano Lett.} \textbf{2016}, \emph{16},
  825--833\relax
\mciteBstWouldAddEndPuncttrue
\mciteSetBstMidEndSepPunct{\mcitedefaultmidpunct}
{\mcitedefaultendpunct}{\mcitedefaultseppunct}\relax
\EndOfBibitem
\bibitem[Pan \latin{et~al.}(2019)Pan, Wang, Zhang, Zhu, Su, Fan, Fu, Huang,
  Wei, Zhang, \latin{et~al.} others]{pan2019dimension}
Pan,~D.; Wang,~J.-Y.; Zhang,~W.; Zhu,~L.; Su,~X.; Fan,~F.; Fu,~Y.; Huang,~S.;
  Wei,~D.; Zhang,~L., \latin{et~al.}  Dimension engineering of high-quality
  InAs nanostructures on a wafer scale. \emph{Nano Lett.} \textbf{2019},
  \emph{19}, 1632--1642\relax
\mciteBstWouldAddEndPuncttrue
\mciteSetBstMidEndSepPunct{\mcitedefaultmidpunct}
{\mcitedefaultendpunct}{\mcitedefaultseppunct}\relax
\EndOfBibitem
\bibitem[Sun \latin{et~al.}(2020)Sun, Gao, Zhang, Yao, Xu, Zheng, Chen, Lu, and
  Zou]{sun2020high}
Sun,~Q.; Gao,~H.; Zhang,~X.; Yao,~X.; Xu,~S.; Zheng,~K.; Chen,~P.; Lu,~W.;
  Zou,~J. High-quality epitaxial wurtzite structured InAs nanosheets grown in
  MBE. \emph{Nanoscale} \textbf{2020}, \emph{12}, 271--276\relax
\mciteBstWouldAddEndPuncttrue
\mciteSetBstMidEndSepPunct{\mcitedefaultmidpunct}
{\mcitedefaultendpunct}{\mcitedefaultseppunct}\relax
\EndOfBibitem
\bibitem[Chen \latin{et~al.}(2021)Chen, Huang, Pan, Xue, Zhang, Zhao, and
  Xu]{chen2021strong}
Chen,~Y.; Huang,~S.; Pan,~D.; Xue,~J.; Zhang,~L.; Zhao,~J.; Xu,~H.~Q. Strong
  and tunable spin-orbit interaction in a single crystalline InSb nanosheet.
  \emph{NPJ 2D Mater. Appl.} \textbf{2021}, \emph{5}, 3\relax
\mciteBstWouldAddEndPuncttrue
\mciteSetBstMidEndSepPunct{\mcitedefaultmidpunct}
{\mcitedefaultendpunct}{\mcitedefaultseppunct}\relax
\EndOfBibitem
\bibitem[Fan \latin{et~al.}(2022)Fan, Chen, Pan, Zhao, and
  Xu]{fan2022electrically}
Fan,~F.; Chen,~Y.; Pan,~D.; Zhao,~J.; Xu,~H.~Q. Electrically tunable
  spin--orbit interaction in an InAs nanosheet. \emph{Nanoscale Advances}
  \textbf{2022}, \emph{4}, 2642--2648\relax
\mciteBstWouldAddEndPuncttrue
\mciteSetBstMidEndSepPunct{\mcitedefaultmidpunct}
{\mcitedefaultendpunct}{\mcitedefaultseppunct}\relax
\EndOfBibitem
\bibitem[Kang \latin{et~al.}(2018)Kang, Fan, Zhi, Pan, Li, Wang, Guo, Zhao, and
  Xu]{kang2018two}
Kang,~N.; Fan,~D.; Zhi,~J.; Pan,~D.; Li,~S.; Wang,~C.; Guo,~J.; Zhao,~J.;
  Xu,~H.~Q. Two-dimensional quantum transport in free-standing InSb nanosheets.
  \emph{Nano letters} \textbf{2018}, \emph{19}, 561--569\relax
\mciteBstWouldAddEndPuncttrue
\mciteSetBstMidEndSepPunct{\mcitedefaultmidpunct}
{\mcitedefaultendpunct}{\mcitedefaultseppunct}\relax
\EndOfBibitem
\bibitem[Zhi \latin{et~al.}(2019)Zhi, Kang, Li, Fan, Su, Pan, Zhao, Zhao, and
  Xu]{zhi2019supercurrent}
Zhi,~J.; Kang,~N.; Li,~S.; Fan,~D.; Su,~F.; Pan,~D.; Zhao,~S.; Zhao,~J.;
  Xu,~H.~Q. Supercurrent and multiple Andreev reflections in InSb nanosheet SNS
  junctions. \emph{Phys. Status. Solidi (B)} \textbf{2019}, \emph{256},
  1800538\relax
\mciteBstWouldAddEndPuncttrue
\mciteSetBstMidEndSepPunct{\mcitedefaultmidpunct}
{\mcitedefaultendpunct}{\mcitedefaultseppunct}\relax
\EndOfBibitem
\bibitem[Salimian \latin{et~al.}(2021)Salimian, Carrega, Verma, Zannier, Nowak,
  Beltram, Sorba, and Heun]{salimian2021gate}
Salimian,~S.; Carrega,~M.; Verma,~I.; Zannier,~V.; Nowak,~M.~P.; Beltram,~F.;
  Sorba,~L.; Heun,~S. Gate-controlled supercurrent in ballistic InSb nanoflag
  Josephson junctions. \emph{Appl. Phys. Lett.} \textbf{2021}, \emph{119},
  214004\relax
\mciteBstWouldAddEndPuncttrue
\mciteSetBstMidEndSepPunct{\mcitedefaultmidpunct}
{\mcitedefaultendpunct}{\mcitedefaultseppunct}\relax
\EndOfBibitem
\bibitem[Iorio \latin{et~al.}(2023)Iorio, Crippa, Turini, Salimian, Carrega,
  Chirolli, Zannier, Sorba, Strambini, Giazotto, \latin{et~al.}
  others]{iorio2023half}
Iorio,~A.; Crippa,~A.; Turini,~B.; Salimian,~S.; Carrega,~M.; Chirolli,~L.;
  Zannier,~V.; Sorba,~L.; Strambini,~E.; Giazotto,~F., \latin{et~al.}
  Half-integer Shapiro steps in highly transmissive InSb nanoflag Josephson
  junctions. \emph{arXiv:2303.05951} \textbf{2023}, \relax
\mciteBstWouldAddEndPunctfalse
\mciteSetBstMidEndSepPunct{\mcitedefaultmidpunct}
{}{\mcitedefaultseppunct}\relax
\EndOfBibitem
\bibitem[Fan \latin{et~al.}(2020)Fan, Chen, Pan, Zhao, and
  Xu]{fan2020measurements}
Fan,~F.; Chen,~Y.; Pan,~D.; Zhao,~J.; Xu,~H.~Q. Measurements of spin-orbit
  interaction in epitaxially grown InAs nanosheets. \emph{Appl. Phys. Lett.}
  \textbf{2020}, \emph{117}, 132101\relax
\mciteBstWouldAddEndPuncttrue
\mciteSetBstMidEndSepPunct{\mcitedefaultmidpunct}
{\mcitedefaultendpunct}{\mcitedefaultseppunct}\relax
\EndOfBibitem
\bibitem[Doh \latin{et~al.}(2005)Doh, van Dam, Roest, Bakkers, Kouwenhoven, and
  De~Franceschi]{doh2005tunable}
Doh,~Y.-J.; van Dam,~J.~A.; Roest,~A.~L.; Bakkers,~E.~P.; Kouwenhoven,~L.~P.;
  De~Franceschi,~S. Tunable supercurrent through semiconductor nanowires.
  \emph{Science} \textbf{2005}, \emph{309}, 272--275\relax
\mciteBstWouldAddEndPuncttrue
\mciteSetBstMidEndSepPunct{\mcitedefaultmidpunct}
{\mcitedefaultendpunct}{\mcitedefaultseppunct}\relax
\EndOfBibitem
\bibitem[Tinkham(2004)]{tinkham2004introduction}
Tinkham,~M. \emph{Introduction to superconductivity}; Courier Corporation,
  2004\relax
\mciteBstWouldAddEndPuncttrue
\mciteSetBstMidEndSepPunct{\mcitedefaultmidpunct}
{\mcitedefaultendpunct}{\mcitedefaultseppunct}\relax
\EndOfBibitem
\bibitem[Tinkham \latin{et~al.}(2003)Tinkham, Free, Lau, and
  Markovic]{tinkham2003hysteretic}
Tinkham,~M.; Free,~J.; Lau,~C.; Markovic,~N. Hysteretic I- V curves of
  superconducting nanowires. \emph{Phys. Rev. B} \textbf{2003}, \emph{68},
  134515\relax
\mciteBstWouldAddEndPuncttrue
\mciteSetBstMidEndSepPunct{\mcitedefaultmidpunct}
{\mcitedefaultendpunct}{\mcitedefaultseppunct}\relax
\EndOfBibitem
\bibitem[Courtois \latin{et~al.}(2008)Courtois, Meschke, Peltonen, and
  Pekola]{courtois2008origin}
Courtois,~H.; Meschke,~M.; Peltonen,~J.; Pekola,~J.~P. Origin of hysteresis in
  a proximity Josephson junction. \emph{Phys. Rev. Lett.} \textbf{2008},
  \emph{101}, 067002\relax
\mciteBstWouldAddEndPuncttrue
\mciteSetBstMidEndSepPunct{\mcitedefaultmidpunct}
{\mcitedefaultendpunct}{\mcitedefaultseppunct}\relax
\EndOfBibitem
\bibitem[Bratus \latin{et~al.}(1995)Bratus, Shumeiko, and
  Wendin]{bratus1995theory}
Bratus,~E.; Shumeiko,~V.; Wendin,~G. Theory of subharmonic gap structure in
  superconducting mesoscopic tunnel contacts. \emph{Phys. Rev. Lett.}
  \textbf{1995}, \emph{74}, 2110\relax
\mciteBstWouldAddEndPuncttrue
\mciteSetBstMidEndSepPunct{\mcitedefaultmidpunct}
{\mcitedefaultendpunct}{\mcitedefaultseppunct}\relax
\EndOfBibitem
\bibitem[Cuevas \latin{et~al.}(1996)Cuevas, Mart{\'\i}n-Rodero, and
  Yeyati]{cuevas1996hamiltonian}
Cuevas,~J.; Mart{\'\i}n-Rodero,~A.; Yeyati,~A.~L. Hamiltonian approach to the
  transport properties of superconducting quantum point contacts. \emph{Phys.
  Rev. B} \textbf{1996}, \emph{54}, 7366\relax
\mciteBstWouldAddEndPuncttrue
\mciteSetBstMidEndSepPunct{\mcitedefaultmidpunct}
{\mcitedefaultendpunct}{\mcitedefaultseppunct}\relax
\EndOfBibitem
\bibitem[Beenakker(1991)]{beenakker1991universal}
Beenakker,~C. W.~J. Universal limit of critical-current fluctuations in
  mesoscopic Josephson junctions. \emph{Phys. Rev. Lett.} \textbf{1991},
  \emph{67}, 3836\relax
\mciteBstWouldAddEndPuncttrue
\mciteSetBstMidEndSepPunct{\mcitedefaultmidpunct}
{\mcitedefaultendpunct}{\mcitedefaultseppunct}\relax
\EndOfBibitem
\bibitem[Artemenko \latin{et~al.}(1979)Artemenko, Volkov, and
  Zaitsev]{artemenko1979theory}
Artemenko,~S.; Volkov,~A.; Zaitsev,~A. Theory of the nonstationary Josephson
  effect in short superconducting contacts. \emph{Zhurnal Eksperimental'noi i
  Teoreticheskoi Fiziki} \textbf{1979}, \emph{76}, 1816--1833\relax
\mciteBstWouldAddEndPuncttrue
\mciteSetBstMidEndSepPunct{\mcitedefaultmidpunct}
{\mcitedefaultendpunct}{\mcitedefaultseppunct}\relax
\EndOfBibitem
\bibitem[Klapwijk \latin{et~al.}(1982)Klapwijk, Blonder, and
  Tinkham]{klapwijk1982explanation}
Klapwijk,~T.; Blonder,~G.; Tinkham,~M. Explanation of subharmonic energy gap
  structure in superconducting contacts. \emph{Physica B+ C} \textbf{1982},
  \emph{109}, 1657--1664\relax
\mciteBstWouldAddEndPuncttrue
\mciteSetBstMidEndSepPunct{\mcitedefaultmidpunct}
{\mcitedefaultendpunct}{\mcitedefaultseppunct}\relax
\EndOfBibitem
\bibitem[Flensberg \latin{et~al.}(1988)Flensberg, Hansen, and
  Octavio]{flensberg1988subharmonic}
Flensberg,~K.; Hansen,~J.~B.; Octavio,~M. Subharmonic energy-gap structure in
  superconducting weak links. \emph{Phys. Rev. B} \textbf{1988}, \emph{38},
  8707\relax
\mciteBstWouldAddEndPuncttrue
\mciteSetBstMidEndSepPunct{\mcitedefaultmidpunct}
{\mcitedefaultendpunct}{\mcitedefaultseppunct}\relax
\EndOfBibitem
\bibitem[Blonder \latin{et~al.}(1982)Blonder, Tinkham, and
  Klapwijk]{blonder1982transition}
Blonder,~G.; Tinkham,~M.; Klapwijk,~T. Transition from metallic to tunneling
  regimes in superconducting microconstrictions: Excess current, charge
  imbalance, and supercurrent conversion. \emph{Phys. Rev. B} \textbf{1982},
  \emph{25}, 4515\relax
\mciteBstWouldAddEndPuncttrue
\mciteSetBstMidEndSepPunct{\mcitedefaultmidpunct}
{\mcitedefaultendpunct}{\mcitedefaultseppunct}\relax
\EndOfBibitem
\bibitem[Li \latin{et~al.}(2016)Li, Kang, Fan, Wang, Huang, Caroff, and
  Xu]{li2016coherent}
Li,~S.; Kang,~N.; Fan,~D.; Wang,~L.; Huang,~Y.; Caroff,~P.; Xu,~H.~Q. Coherent
  charge transport in ballistic InSb nanowire Josephson junctions. \emph{Sci.
  Rep.} \textbf{2016}, \emph{6}, 24822\relax
\mciteBstWouldAddEndPuncttrue
\mciteSetBstMidEndSepPunct{\mcitedefaultmidpunct}
{\mcitedefaultendpunct}{\mcitedefaultseppunct}\relax
\EndOfBibitem
\bibitem[Nishio \latin{et~al.}(2011)Nishio, Kozakai, Amaha, Larsson, Nilsson,
  Xu, Zhang, Tateno, Takayanagi, and Ishibashi]{nishio2011supercurrent}
Nishio,~T.; Kozakai,~T.; Amaha,~S.; Larsson,~M.; Nilsson,~H.~A.; Xu,~H.~Q.;
  Zhang,~G.; Tateno,~K.; Takayanagi,~H.; Ishibashi,~K. Supercurrent through
  inas nanowires with highly transparent superconducting contacts.
  \emph{Nanotechnol.} \textbf{2011}, \emph{22}, 445701\relax
\mciteBstWouldAddEndPuncttrue
\mciteSetBstMidEndSepPunct{\mcitedefaultmidpunct}
{\mcitedefaultendpunct}{\mcitedefaultseppunct}\relax
\EndOfBibitem
\bibitem[Dinsmore~III \latin{et~al.}(2008)Dinsmore~III, Bae, and
  Bezryadin]{dinsmore2008fractional}
Dinsmore~III,~R.~C.; Bae,~M.-H.; Bezryadin,~A. Fractional order Shapiro steps
  in superconducting nanowires. \emph{Appl. Phys. Lett.} \textbf{2008},
  \emph{93}, 192505\relax
\mciteBstWouldAddEndPuncttrue
\mciteSetBstMidEndSepPunct{\mcitedefaultmidpunct}
{\mcitedefaultendpunct}{\mcitedefaultseppunct}\relax
\EndOfBibitem
\bibitem[Raes \latin{et~al.}(2020)Raes, Tubsrinuan, Sreedhar, Guala, Panghotra,
  Dausy, de~Souza~Silva, and Van~de Vondel]{raes2020fractional}
Raes,~B.; Tubsrinuan,~N.; Sreedhar,~R.; Guala,~D.; Panghotra,~R.; Dausy,~H.;
  de~Souza~Silva,~C.~C.; Van~de Vondel,~J. Fractional Shapiro steps in
  resistively shunted Josephson junctions as a fingerprint of a skewed
  current-phase relationship. \emph{Phys. Rev. B} \textbf{2020}, \emph{102},
  054507\relax
\mciteBstWouldAddEndPuncttrue
\mciteSetBstMidEndSepPunct{\mcitedefaultmidpunct}
{\mcitedefaultendpunct}{\mcitedefaultseppunct}\relax
\EndOfBibitem
\bibitem[Rokhinson \latin{et~al.}(2012)Rokhinson, Liu, and
  Furdyna]{rokhinson2012fractional}
Rokhinson,~L.~P.; Liu,~X.; Furdyna,~J.~K. The fractional ac Josephson effect in
  a semiconductor--superconductor nanowire as a signature of Majorana
  particles. \emph{Nat. Phys.} \textbf{2012}, \emph{8}, 795--799\relax
\mciteBstWouldAddEndPuncttrue
\mciteSetBstMidEndSepPunct{\mcitedefaultmidpunct}
{\mcitedefaultendpunct}{\mcitedefaultseppunct}\relax
\EndOfBibitem
\bibitem[Wiedenmann \latin{et~al.}(2016)Wiedenmann, Bocquillon, Deacon,
  Hartinger, Herrmann, Klapwijk, Maier, Ames, Br{\"u}ne, Gould, \latin{et~al.}
  others]{wiedenmann20164}
Wiedenmann,~J.; Bocquillon,~E.; Deacon,~R.~S.; Hartinger,~S.; Herrmann,~O.;
  Klapwijk,~T.~M.; Maier,~L.; Ames,~C.; Br{\"u}ne,~C.; Gould,~C.,
  \latin{et~al.}  4 $\pi$-periodic Josephson supercurrent in HgTe-based
  topological Josephson junctions. \emph{Nat. Commun.} \textbf{2016}, \emph{7},
  10303\relax
\mciteBstWouldAddEndPuncttrue
\mciteSetBstMidEndSepPunct{\mcitedefaultmidpunct}
{\mcitedefaultendpunct}{\mcitedefaultseppunct}\relax
\EndOfBibitem
\bibitem[Rosenbach \latin{et~al.}(2021)Rosenbach, Schmitt, Sch{\"u}ffelgen,
  Stehno, Li, Schleenvoigt, Jalil, Mussler, Neumann, Trellenkamp,
  \latin{et~al.} others]{rosenbach2021reappearance}
Rosenbach,~D.; Schmitt,~T.~W.; Sch{\"u}ffelgen,~P.; Stehno,~M.~P.; Li,~C.;
  Schleenvoigt,~M.; Jalil,~A.~R.; Mussler,~G.; Neumann,~E.; Trellenkamp,~S.,
  \latin{et~al.}  Reappearance of first Shapiro step in narrow topological
  Josephson junctions. \emph{Sci. Adv.} \textbf{2021}, \emph{7}, eabf1854\relax
\mciteBstWouldAddEndPuncttrue
\mciteSetBstMidEndSepPunct{\mcitedefaultmidpunct}
{\mcitedefaultendpunct}{\mcitedefaultseppunct}\relax
\EndOfBibitem
\end{mcitethebibliography}

\newpage

\begin{figure*}[!h] 
\centering
\includegraphics[width=0.5\linewidth] {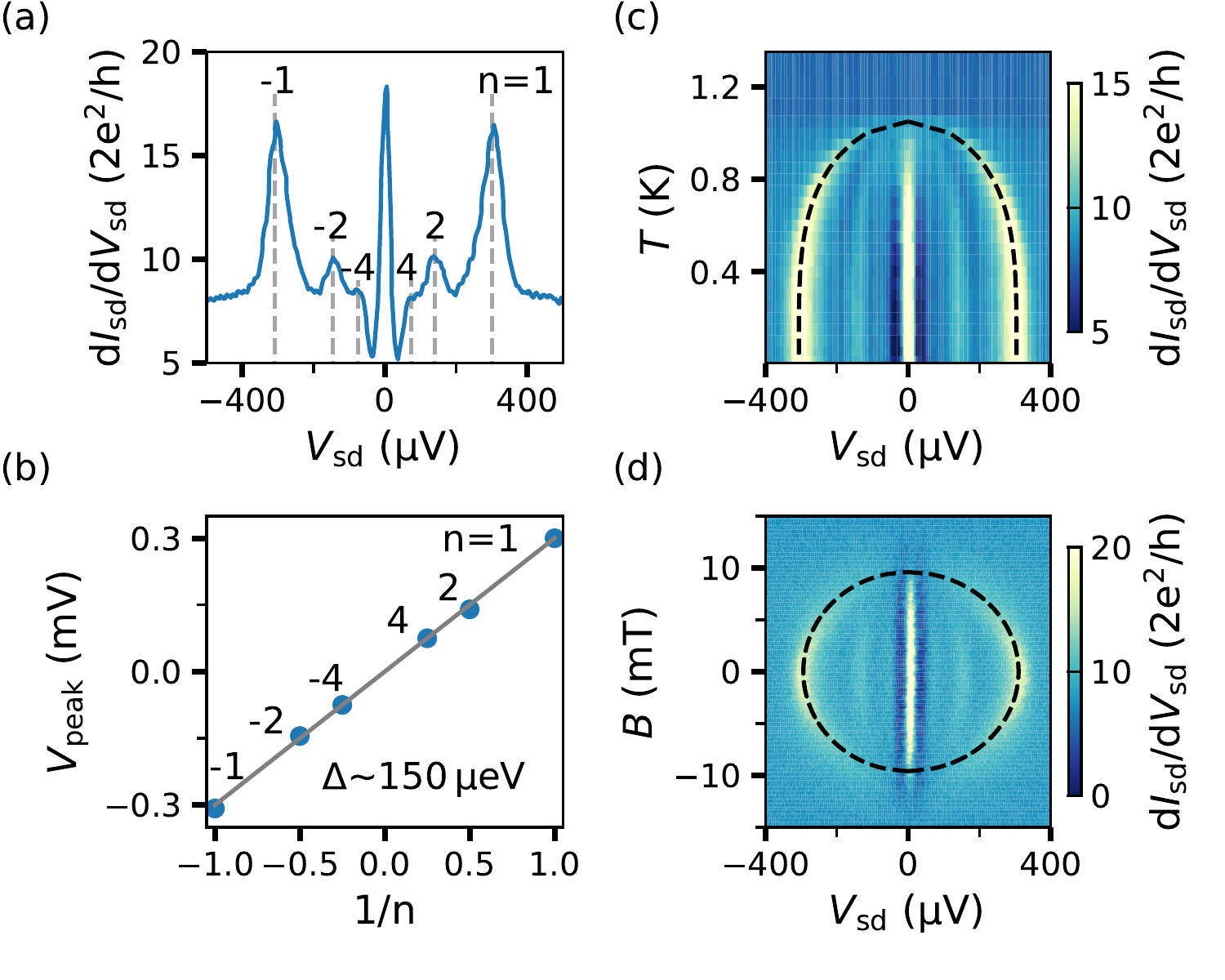}
\caption{\doublespacing \textbf{(a)} Differential conductance $dI_{\mathrm{sd}}$/$dV_{\mathrm{sd}}$ as a function of source drain voltage $V_{\mathrm{sd}}$ measured at $V_{\mathrm{bg}}=0$, $B=0$ and base temperature $T\sim 20\,\mathrm{mK}$. The vertical grey dashed lines mark the peak positions of MAR peaks, and the integer numbers $n$ represent the orders of the MARs. \textbf{(b)} Plot of the MAR peak position ($V_\mathrm{peak}$) against the inverse MAR order $1/n$. The grey line is the linear fit of data. The superconducting energy gap $\Delta$ determined from the slope of the fitting line is $\Delta \sim 150\,\mathrm{\mu eV}$. \textbf{(c)} Differential conductance $dI_{\mathrm{sd}}$/$dV_{\mathrm{sd}}$ measured as a function of $V_{\mathrm{sd}}$ and $T$ at $V_{\mathrm{bg}}=0$ and $B=0$. The black dashed line is the fitting curve of the first-order MAR peak positions, $V_{\mathrm{sd}}=\pm\,2\Delta/e$, to the BCS theory. \textbf{(d)} Differential conductance $dI_{\mathrm{sd}}$/$dV_{\mathrm{sd}}$ measured as a function of $V_{\mathrm{sd}}$ and magnetic field $B$ at $V_{\mathrm{bg}}=0$ and $T\sim 20$ mK. The black dashed line is the fitting curve of the first-order MAR positions, $V_{\mathrm{sd}}=\pm\,2\Delta/e$, to the BCS theory.}\label{figure:3}
\end{figure*}

\end{document}


\section{Methodes}
\noindent\textbf{Material growth} 
\par
\begin{figure*}[!h] 
\centering
\includegraphics[width=1\linewidth]{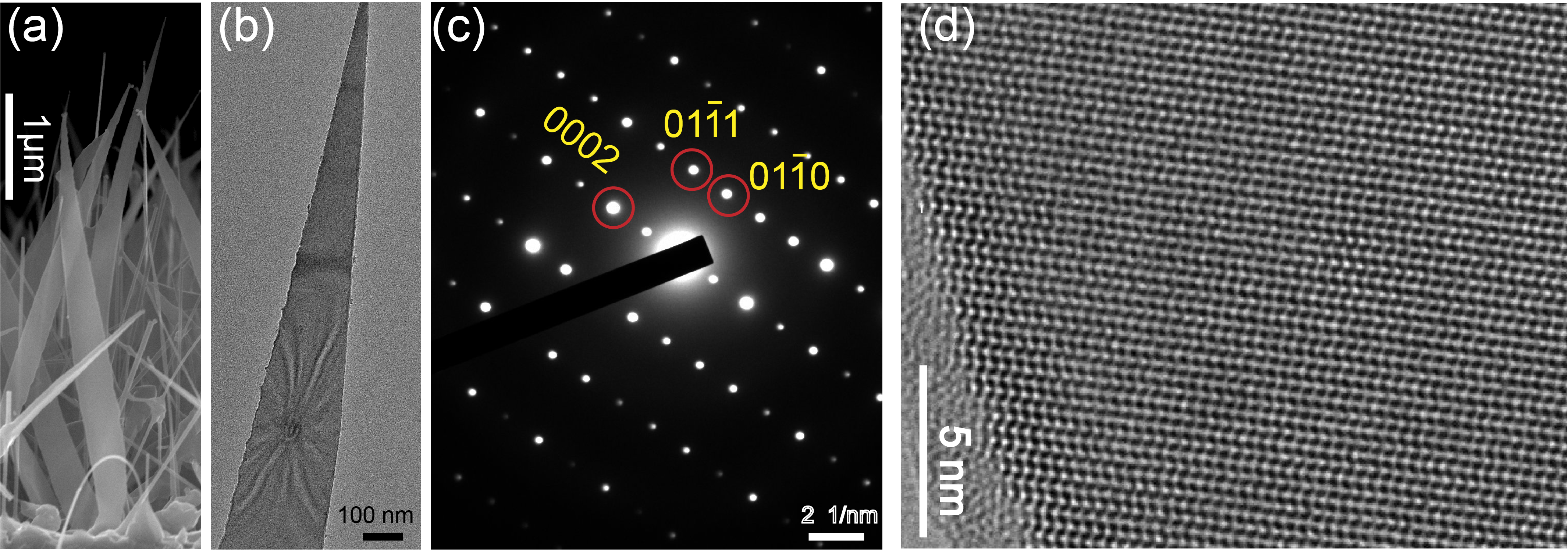}
\caption{\doublespacing \textbf{(a)} SEM image (side view) of InAs nanosheets on the grown substrate. \textbf{(b)} Overview TEM image of a typical InAs nanosheet. \textbf{(c)} SAED pattern of the InAs nanosheet. \textbf{(d)} High resolution TEM image of the InAs nanosheet. } \label{figure:S1}
\end{figure*}

 Free-standing InAs nanosheets are grown on a p-type Si(111) substate by MBE at a temperature of 545 $^{\circ}$C, see Ref. 29 in the main article for detailed information. Figure \ref{figure:S1}(a) shows a scanning electron microscope (SEM) image (side view) of InAs nanosheets on the growth substrate. Figure \ref{figure:S1}(b) shows a transmission electron microscopy (TEM) image of a typical InAs nanosheet. Figure \ref{figure:S1}(c) displays a corresponding selected-area electron diffraction (SAED) pattern recorded along the [2-1-10] crystallographic direction of the InAs nanosheet. Figure \ref{figure:S1}(d) shows a high-resolution TEM image of the nanosheet, which verifies the high quality of the crystal with nearly no stacking faults and twin defects. 

\noindent\textbf{Device fabrication} 
 \par InAs nanosheets are mechanically transferred onto a degenerately n-doped  $\mathrm{Si}$ substrate, capped with a $300\,\mathrm{nm}$ thick layer of ${\mathrm{SiO_2}}$ on top, with pre-patterned Ti/Au markers on the surface. The highly doped Si and the $300\,\mathrm{nm}$ thick ${\mathrm{SiO_2}}$ are used as a global back gate and the gate dielectric, respectively. SEM is used to select suitable InAs nanosheets for device fabrication. PMMA A4 with a thickness of about $200\,\mathrm{nm}$ is spin coated on the surface and then baked at a temperature of $170^\text{o}$C for 3 minutes. Standard electron-beam lithography is used to pattern the device contact areas on the PMMA. The patterned sample is then developed in  a standard development process in developer $\mathrm{MIBK:IPA}=1:3$ at room temperature. In-situ argon ion milling is used to remove the oxide layer on the opened surface of the InAs nanosheets right before deposition of $5\,\mathrm{nm}/80\,\mathrm{nm}$ of Ti/Al by e-beam evaporation. The sample is then bathed in acetone for lift off and cleaned by rinsing in isopropyl alcohol. A room temperature probe station is employed to measure the resistance of the devices fabricated on the chip and devices with a resistance of about $1\sim2\,\mathrm{k\Omega}$ are selected for low temperature transport measurements. A SEM image of a typical device, which is also the device with the measurements reported in the  main article, is shown in Figure 1(a) of the main article. Note that the SEM image of the device is taken taken after all the low temperature transport measurements are made.
 
 \noindent\textbf{Low temperature transport measurements} 
 \par Devices are measured in a dilution refrigerator equipped with a magnet. Two different measurement configurations are used in this work. The first one is  a dc current bias configuration, see Figure S2(a). In this configuration, a quasi-four-terminal measurement setup is employed. The source-drain current is pre-amplified by $10^6 \,\mathrm{A/V}$ and the measured voltage is pre-amplified by setting a gain at $100$ [with an actual gain at $\sim 117$, see Figure \ref{figure:S3}(a) and Figure \ref{figure:S3}(b)]. The second one is a voltage bias configuration, see Figure S2(b). In this configuration, a  two-terminal measurement setup is employed and the measurements are performed using a standard lock-in technique by applying a (dc+ac)  bias voltage to the source contact, see Figure S2(b). Here, a serial resistance of $422\,\mathrm{\Omega}$ from the measurement circuit has been subtracted. Figure \ref{figure:S3}(c) and Figure \ref{figure:S3}(d) show the circuit for the measurement of the serial resistance of the measurement circuit and how the serial resistance value is extracted. The value of the magnetic field is calibrated by an offset of $-6\,\mathrm{mT}$.

\begin{figure*}[!h] 
\centering
\includegraphics[width=1\linewidth]{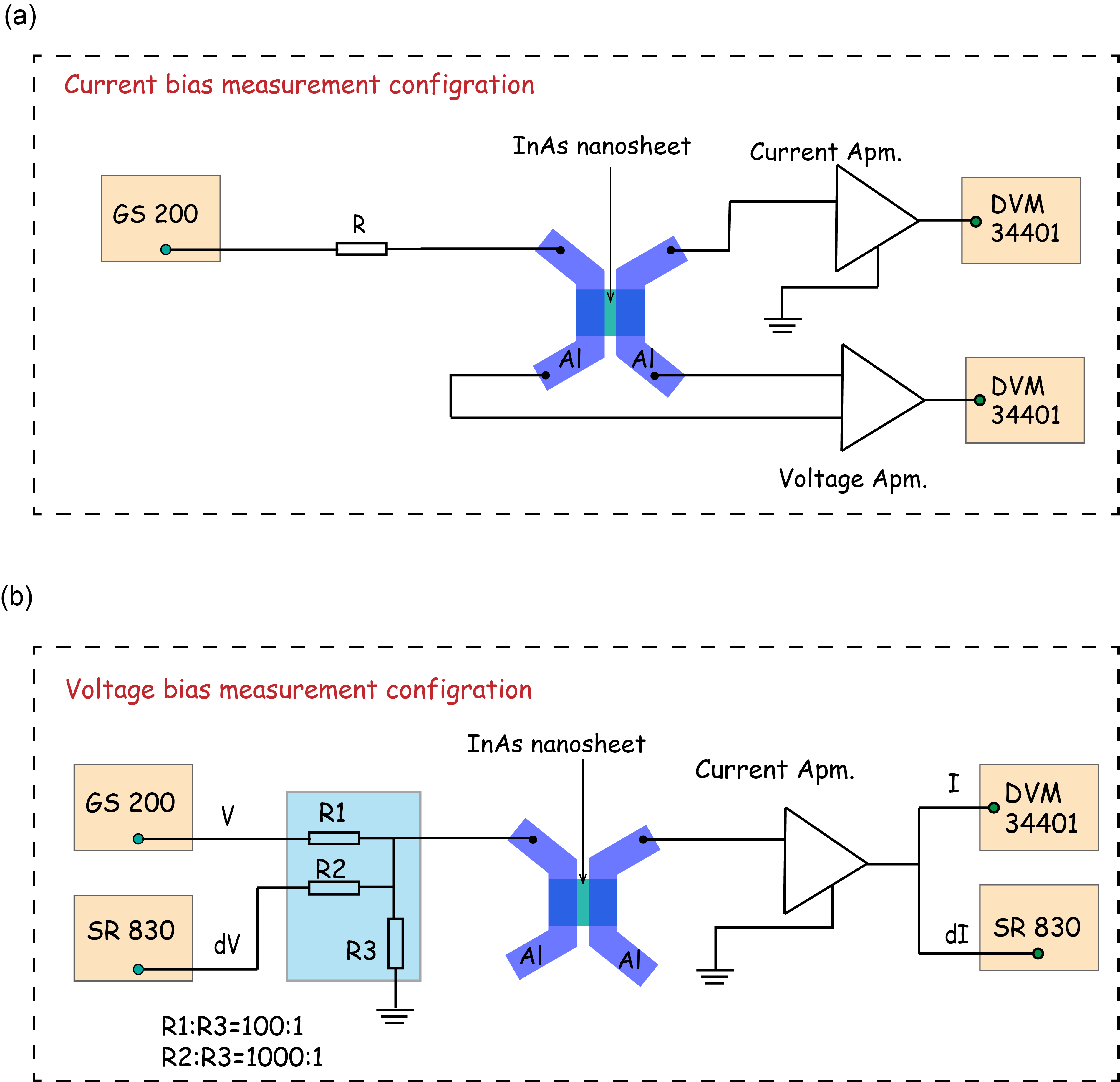}
\caption{\doublespacing \textbf{(a)} Measurement circuit setup for the current bias configuration. \textbf{(b)} Measurement circuit setup for the voltage bias configuration. } \label{figure:S2}
\end{figure*}
 
\begin{figure*}[!h]
\centering
\includegraphics[width=1\linewidth]{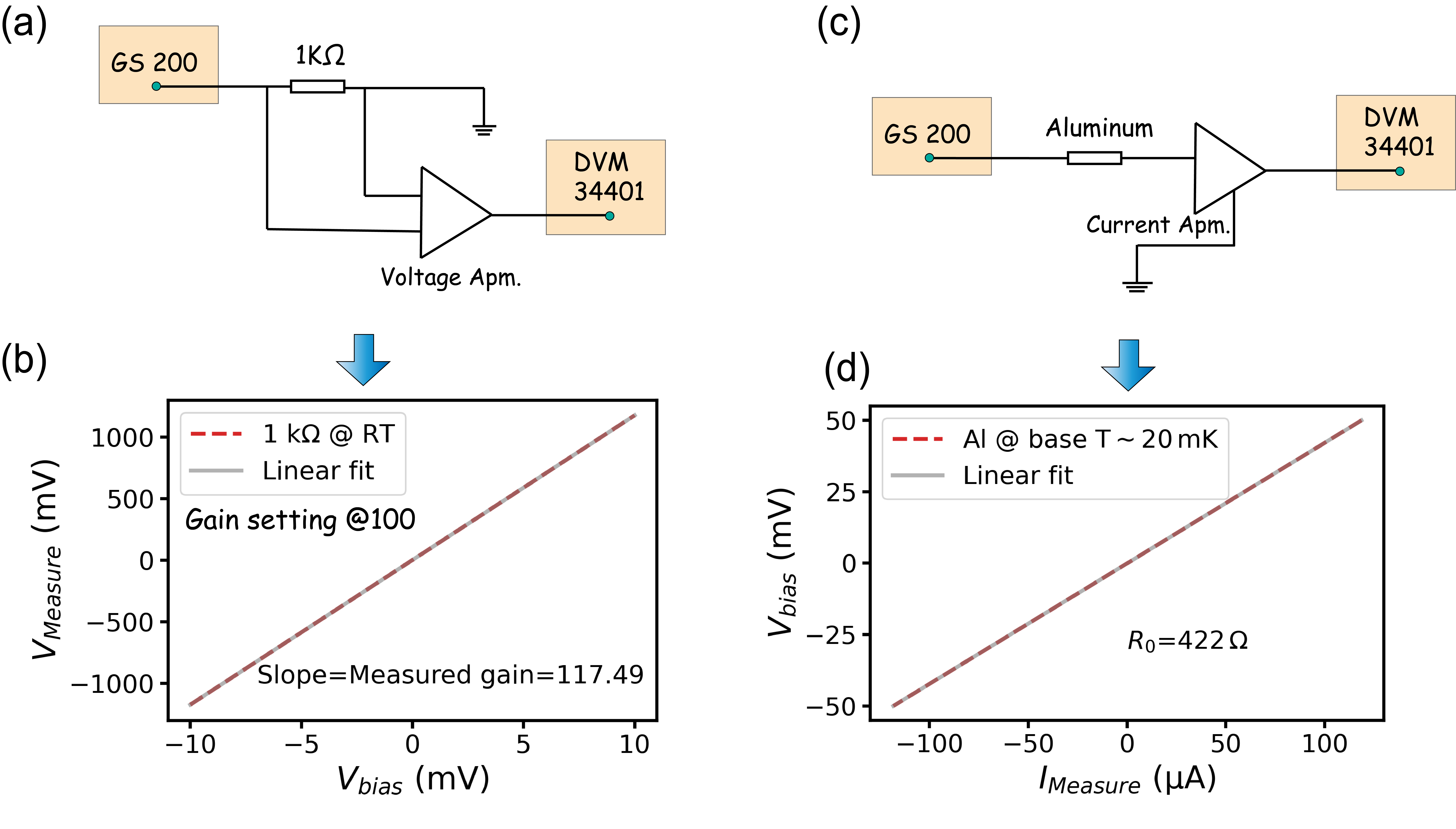} 
\caption{\doublespacing \textbf{(a)} Measurement circuit setup for calibration of the voltage amplifier. \textbf{(b)} Measured voltage $V_\mathrm{Measure}$ with a voltage pre-amplifier (with a gain set at 100) as a function of bias voltage $V_\mathrm{bias}$ (red dashed line). The slope of the linear fit represents the actual gain of the voltage pre-amplifier at a setting gain of 100, which is 117. \textbf{(c)} Measurement circuit setup for calibrating the serial resistance in the measurement circuit. \textbf{(d)} Plot of voltage bias $V_\mathrm{bias}$ versus measured current $I_\mathrm{Measure}$ of an aluminum strip with a two-terminal configuration at a temperature of $\sim 20\,\mathrm{mK}$. The resistance of the aluminum strip is zero for the fact that it is at the superconducting state. Thus, the slope of the linear fit of the measured curve corresponds to the serial resistance in the measurement circuit, which is $422\,\mathrm{\Omega}$.} \label{figure:S3}
\end{figure*}

\clearpage

\section{Additional data}
\par The measurement data presented in the main article all come from the InAs nanosheet Josephson junction device shown in Figure 1(a) of the main article (denoted as device D1). Here we display the measurement data from the second device (denoted as device D2) as well as some additional measurement data from device D1. Devices D2 and D1 are fabricated at the same time on the same substrate with the same fabrication process. 
\begin{figure*}[!h]
\centering
\includegraphics[width=0.75\linewidth]{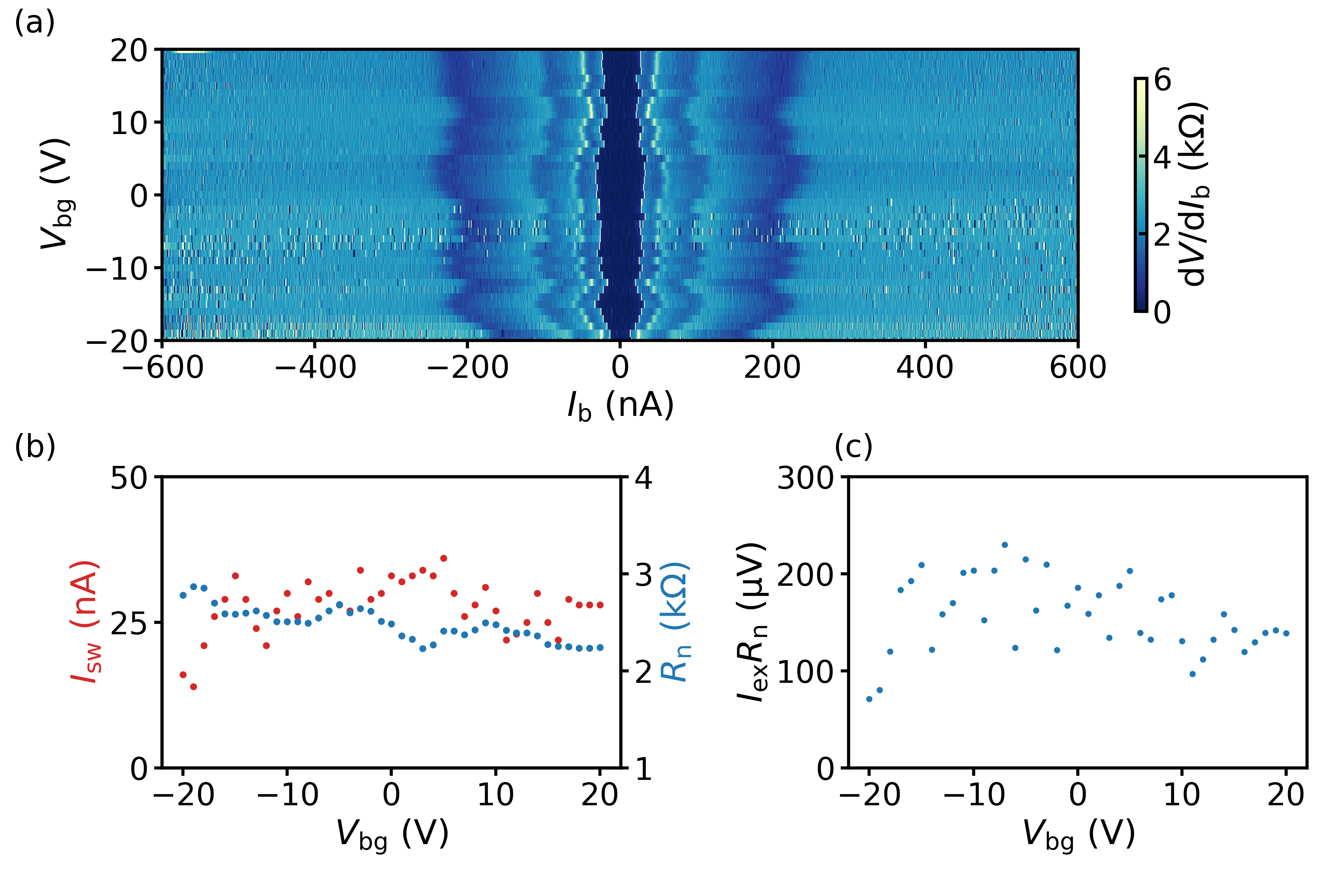} 
\caption{\doublespacing \textbf{Gate dependent supercurrent measurements of device D2.} \textbf {(a)} Differential resistance $dV$/$dI_{\mathrm{b}}$ as a function of bias current $I_{\mathrm{b}}$ and back gate voltage $V_{\mathrm{bg}}$ (upward current sweeping direction only). The central dark area is the superconducting region with $dV$/$I_{\mathrm{b}} =0$. The switching current $I_{\mathrm{sw}}$ can be extracted from the right edge of the centre dark area, which varies with $V_{\mathrm{bg}}$. \textbf{(b)} $I_{\mathrm{sw}}$ and normal state resistance $R_{\mathrm{n}}$ extracted from (a) as a function of $V_{\mathrm{bg}}$. The value of the $I_{\mathrm{sw}}R_{\mathrm{n}}$ product fluctuates in a range between $40\,\mathrm{\mu eV}$ and $90\,\mathrm{\mu eV}$ as $V_{\mathrm{bg}}$ varies from $-20\,\mathrm{V}$ to $20\,\mathrm{V}$. \textbf{(c)} $I_{\mathrm{ex}}R_{\mathrm{n}}$ product as a function of $V_\mathrm{bg}$, where $I_{\mathrm{ex}}$ is the excess current extracted from (a). The value of $I_\mathrm{ex}R_{\mathrm{n}}$ varies between $70\,\mathrm{\mu eV}$ and $230\,\mathrm{\mu eV}$, leading to that the interface transparency of the device varies in a range between $60\,\%$ and $86\,\%$. All data are measured at $B=0$ and $T\sim 20\,\mathrm{mK}$.}\label{figure:S4}
\end{figure*}

\begin{figure*}[!h] 
\centering
\includegraphics[width=1\linewidth]{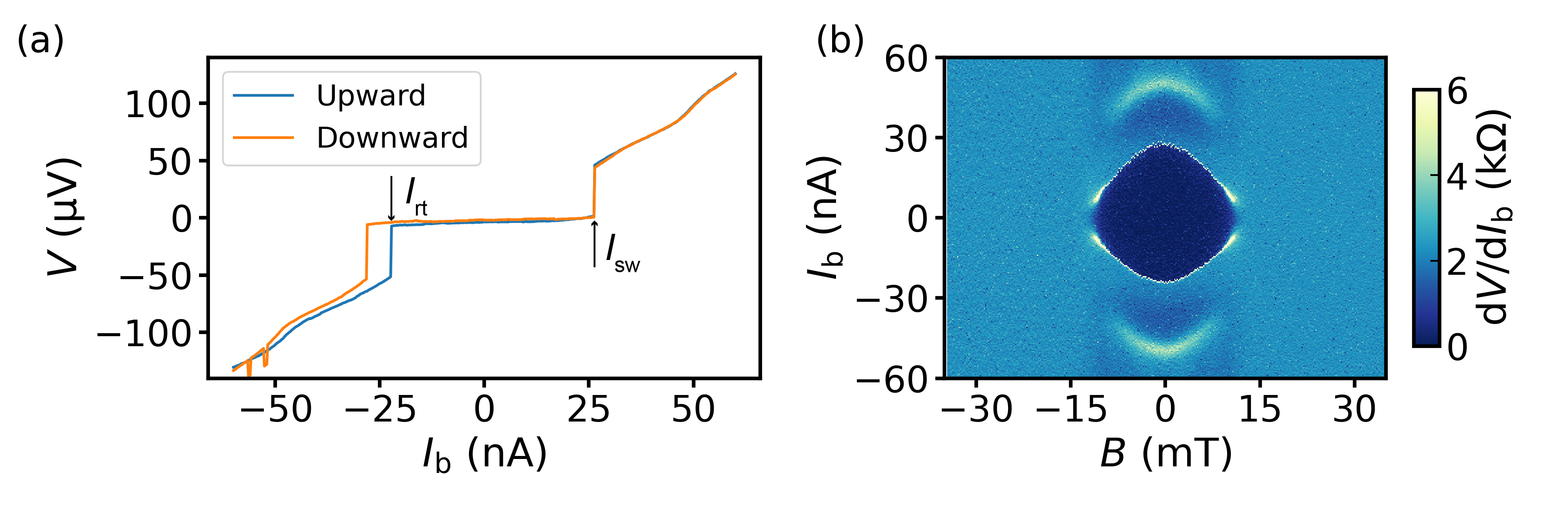}
\caption{\doublespacing \textbf{Hysteresis behavior and magnetic field dependent measurements of device D2 in the current bias configuration.} \textbf {(a)} Voltage $V$ across the junction as a function of bias current $I_{\mathrm{b}}$ measured for both upward and downward current sweep directions. The data show a similar hysteresis behaviour as device D1. \textbf{(b)} Differential resistance $dV$/$dI_{\mathrm{b}}$ as a function of bias current $I_{\mathrm{b}}$ and magnetic field $B$. The central dark area corresponds to the region where the junction is at the superconducting state with $dV$/$dI_{\mathrm{b}}=0$. The upper and lower fringes are due to multiple Andreev reflections. Here, the Same as device D1, no side lobes of the Fraunhofer pattern is observed.}\label{figure:S5}
\end{figure*}

\begin{figure*}[!h] 
\centering
\includegraphics[width=1\linewidth]{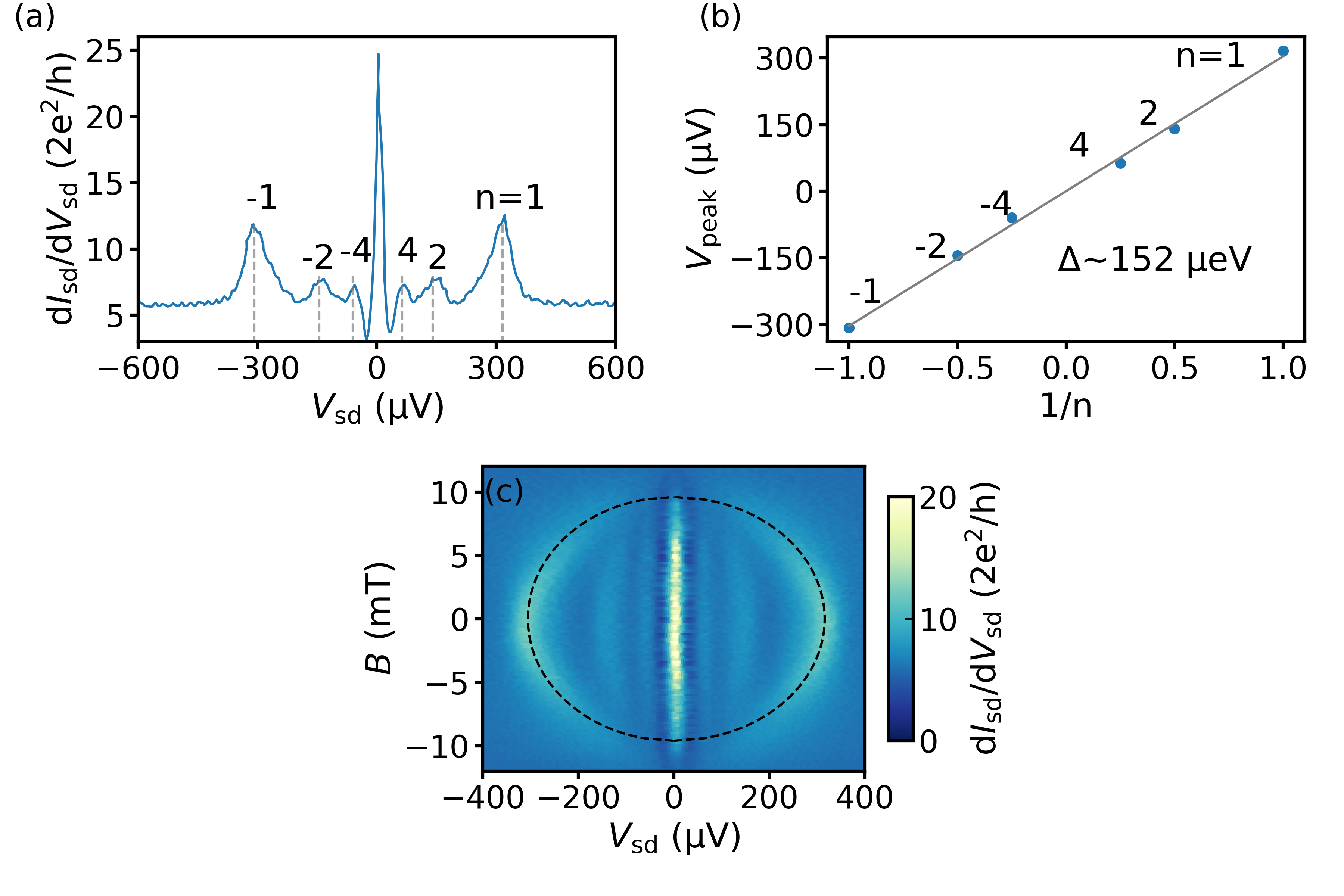}
\caption{\doublespacing \textbf{Measurements of device D2 in the voltage bias configuration.} 
\textbf {(a)} Differential conductance $dI_{\mathrm{sd}}$/$dV_{\mathrm{sd}}$ as a function of source drain voltage $V_{\mathrm{sd}}$ at base temperature $T\sim 20\,\mathrm{mK}$, $B=0$ and $V_{\mathrm{bg}}=0$. The vertical grey dashed lines mark the positions of multiple Andreev reflection peaks and the integer numbers $n$ denote the orders of the multiple Andreev reflection peaks. \textbf{(b)} Plot of the peak positions ($V_\mathrm{peak}$) versus the inverse orders $1/n$ of the multiple Andreev reflections. The grey line is the linear fit of the data points. The superconducting energy gap determined from the slope of the fitting line is $\sim 152\,\mathrm{\mu eV}$. \textbf{(c)} Differential conductance $dI_{\mathrm{sd}}$/$dV_{\mathrm{sd}}$, on a color scale, measured as a function of $V_{\mathrm{sd}}$ and $B$. The black dashed line is the fitting curve of the magnetic field dependence of the peak position of the first order multiple Andreev reflections, $V_{\mathrm{sd}}=\pm\,2\Delta/e$, according to $\Delta(B)={\Delta(0)}[1-(B/B_c)^2]^{1/2}$. The extracted $\Delta(0)$ and $B_{\mathrm{c}}$ from the fit are $\sim 155\,\mathrm{\mu eV}$ and $\sim 9.6\,\mathrm{mT}$, respectively.}\label{figure:S6}
\end{figure*}

\begin{figure*}[!h] 
\centering
\includegraphics[width=1\linewidth]{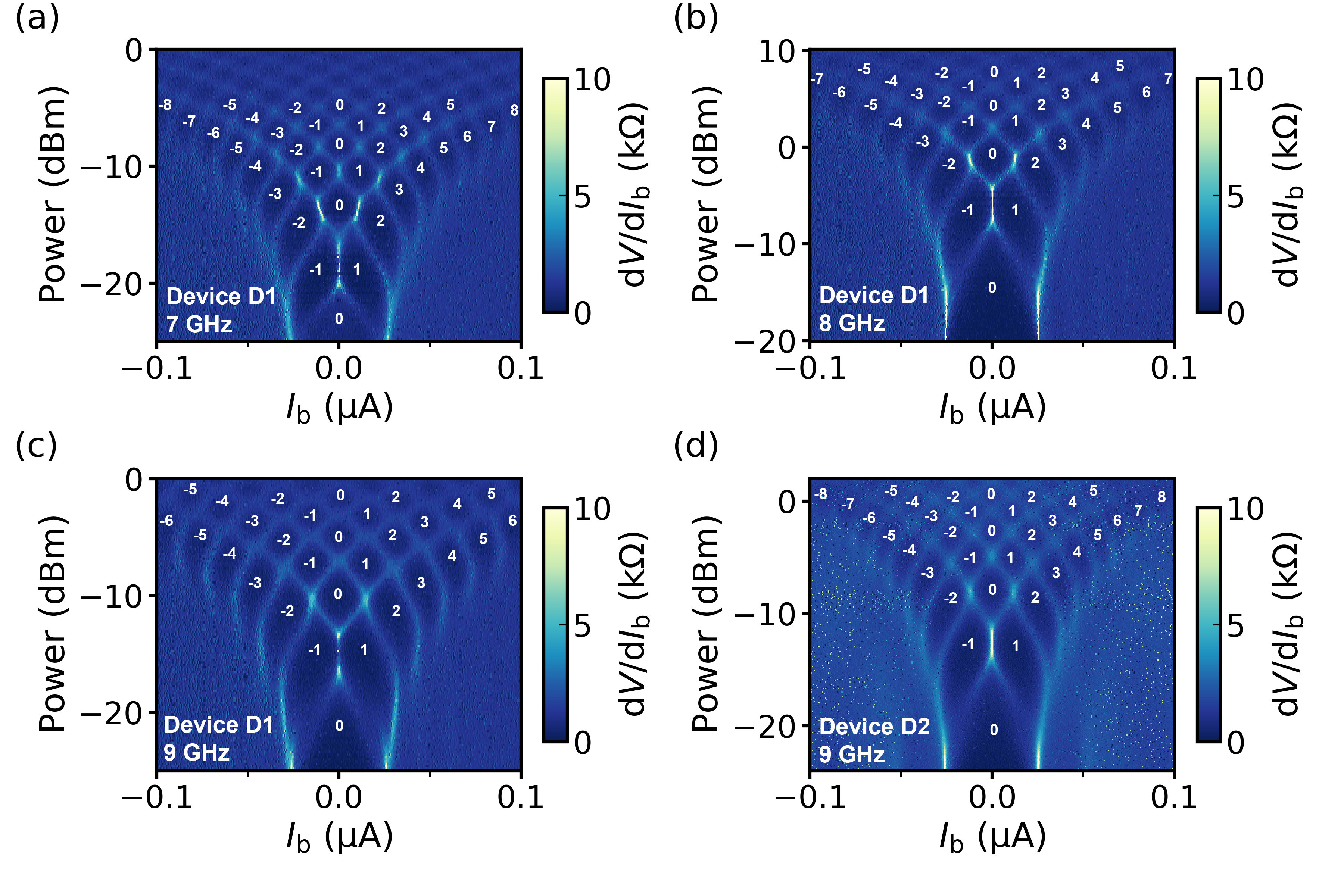}
\caption{\doublespacing \textbf{Additional data of the measurements of device D1 under microwave radiation and data of the measurements of device D2 under microwave radiation.} \textbf{(a)}, \textbf{(b)} and \textbf{(c)} Differential resistance measured for device D1 under microwave radiation as a function of bias current $I_{\mathrm{b}}$ and microwave power at microwave frequencies of $f=7\,\mathrm{GHz}$, $f=8\,\mathrm{GHz}$ and $f=9\,\mathrm{GHz}$, respectively, and at $V_{\mathrm{bg}}=0\,\mathrm{V}$, $B=0$ and $T\sim 20\,\mathrm{mK}$. The dark blue areas marked by integer numbers $n$ are the regions where the Shapiro steps of orders $n$ are observed.  \textbf{(d)} The same as in panels (a) to (c), but for the measurements of device D2 under microwave radiation at frequency $f=9\,\mathrm{GHz}$.}\label{figure:S7}
\end{figure*}